\documentclass[nonblindrev]{informs3-rev}

\OneAndAHalfSpacedXI 

\usepackage{endnotes}
\let\footnote=\endnote

  \usepackage{amssymb}
\usepackage{graphicx}
\usepackage{amsmath}
\usepackage{bm} 
\usepackage{hyperref}
\usepackage{color}
\usepackage{dsfont}

\newcommand{\id}{\mathds{1}}

\newcommand{\E}{\mathbb{E}}
\newcommand{\X}{\mathcal{X}}
\newcommand{\N}{\mathbb{N}}

\usepackage{tikz} 
\usepackage{caption} 
\usetikzlibrary{decorations.pathreplacing}

 \usepackage{pgfplots}
 \usetikzlibrary{calc,patterns}

\TheoremsNumberedThrough     
\ECRepeatTheorems

\EquationsNumberedThrough    

\MANUSCRIPTNO{(Preprint)} 

\begin{document}


\RUNAUTHOR{Castagnoli et al.}

\RUNTITLE{Star-shaped Risk Measures}

\TITLE{Star-shaped Risk Measures}

\ARTICLEAUTHORS{%
\AUTHOR{Erio Castagnoli}
\AFF{Department of Decision Sciences, Bocconi University, MI, Italy, \EMAIL{tributoalduca@gmail.com}} 
\AUTHOR{Giacomo Cattelan}
\AFF{Department of Economics, New York University, NY, USA, \EMAIL{gc2507@nyu.edu}}
\AUTHOR{Fabio Maccheroni}
\AFF{Department of Decision Sciences, Bocconi University, MI, Italy, \EMAIL{fabio.maccheroni@unibocconi.it}}
\AUTHOR{Claudio Tebaldi}
\AFF{Department of Finance, Bocconi University, MI, Italy, \EMAIL{claudio.tebaldi@unibocconi.it}}
\AUTHOR{Ruodu Wang}
\AFF{Department of Statistics and Actuarial Science, University of Waterloo, ON, Canada, \EMAIL{wang@uwaterloo.ca}}

} 

\ABSTRACT{%
In this paper monetary risk measures that are positively superhomogeneous,
called \emph{star-shaped risk measures}, are characterized and their
properties are studied. The measures in this class, which arise when the subadditivity property of coherent risk measures is dispensed
with and positive homogeneity is weakened, include
all practically used risk measures, in particular,
 both convex risk measures
and Value-at-Risk. 
From a financial viewpoint, our relaxation of convexity is necessary to
quantify the capital requirements for risk exposure in the presence of liquidity risk, 
competitive delegation, or robust aggregation mechanisms. 
From a decision theoretical perspective, star-shaped risk measures emerge
from variational preferences when risk mitigation strategies can be adopted
by a rational decision maker. 
}%



\maketitle

%


\section{Introduction} 
 
The calculation of regulatory capital requirements  is critical for the
well-functioning of the financial sector.
On this important issue,
Artzner et al.~\cite{ADEH99} redefined the way of thinking about risk
measurement.  In their language, a risk measure $\rho $ associates to every
uncertain future loss $X$ a capital requirement $\rho  ( X ) $. 
This quantity is interpreted as the minimum amount of money a
financial institution, which holds a position described by $X$, must have in
order to absorb the realized losses and keep solvent in good times and bad.
For instance,
Value-at-Risk (VaR) at level 99\% 
is the minimum amount of money  that shields against insolvency (losses
greater than reserves) in at least the $99\%$ of cases.

A \emph{coherent risk measure} of
Artzner et al.~\cite{ADEH99} is a risk measure which satisfies monotonicity, translation invariance, normalization, positive homogeneity and subadditivity (to be defined formally in Section \ref{sec:def}).
All of these properties are motivated by the interpretation of $\rho (X)$ as
a buffer against future losses, possibly imposed on a financial institution
by an (internal or external) regulator. The \emph{monotonicity} property,
just says that higher losses require higher reserves. In the presence of a
market of riskless bonds, \emph{translation invariance} establishes that the
acquisition of a claim with sure payoff  reduces by the same
amount the risk of the position previously held.
The meaning of \emph{%
normalization},  i.e., $\rho (0)=0$, is self-evident (and it is implied by the other properties). 
These first three properties are broadly accepted, and we assume them
throughout when we speak of \emph{risk measures}.\footnote{\emph{For the
connoisseur.} The complete name would be \textquotedblleft 
 monetary  and normalized risk measures.\textquotedblright\
Of course there are meaningful \textquotedblleft risk
measures\textquotedblright\ that do not satisfy these properties, see, e.g., Cerreia-Vioglio et al. \cite{CMMM11} and
Farkas, Koch-Medina, and Munari \cite{FMM14}, but their study goes beyond
the scope of this paper.}

On the other hand, since the very beginning, \emph{positive homogeneity} has
been object of concerns, and it actually motivated the introduction of
convex risk measures by F\"{o}llmer and Schied \cite[p. 430]{FS02} and
Frittelli and Rosazza Gianin \cite[p.~1475]{FR02}. To justify  
positive homogeneity, Artzner et al.~\cite[p.~209]{ADEH99} argue as follows: for each
risk $X$, it must be the case that%
\begin{equation}
\rho (\lambda X)\geq \lambda \rho (X)\qquad \forall \lambda >1 
\tag{$\bigstar$} \label{eq:star}
\end{equation}%
because additional liquidity risk may arise if a position is multiplied by a
large factor.  The \textquotedblleft star
shape\textquotedblright\ property \eqref{eq:star} has a simple and straight economic motivation, 
but under  subadditivity, it is equivalent to
positive homogeneity.
As such, subadditivity tends to ignore concentration effects by forcing the inequality in  \eqref{eq:star} to be an equality, which may be an unintended consequence.
This emphasizes the problematic nature of {subadditivity} in {some} specific contexts. Section \ref{sec:ECB} presents several  examples where a strict inequality in \eqref{eq:star} is compelling. 

%
%
%
%

The present paper studies the theory of risk measures when subadditivity is dispensed with,
and positive homogeneity replaced with the genuinely weaker star shapedness. 
Throughout the paper, a map $\rho $ is 
\emph{star-shaped} if it satisfies condition \eqref{eq:star}.

As noted by both Frittelli and Rosazza Gianin \cite{FR02} and F\"{o}llmer and Schied \cite{FS02},
convex risk measures satisfy this property, which might be called
\textquotedblleft convexity at $0$\textquotedblright\ and has also been called
\textquotedblleft positive superhomogeneity\textquotedblright\ for  
obvious mathematical reasons. Like convexity, star shapedness penalizes
concentration of risk and the ensuing liquidity problems (the risk-to-exposure ratio of any
position is weakly increasing; see Proposition \ref{pro:star_char}), but
differently from convexity, star shapedness does not take stance on the
effects of \textquotedblleft merge-and-downsize\textquotedblright\
strategies. 
More precisely, star shapedness 
reflects the idea that
downsizing reduces risk more than proportionally, and on top of that, convexity further requires %
that the risk-diminishing effects of downsizing more than compensate any
possible additional risk entailed by a merger.

Our choice of the name \textquotedblleft star-shaped\textquotedblright\ over
its competitors, is motivated by the fact that a risk measure is star-shaped
if, and only if, the set of acceptable risky positions it determines ---that
is, the set of positions that do not require any accumulation of capital---
is star-shaped.  
The meaning is
transparent: reducing the exposure in an acceptable position cannot make it
unacceptable. The implications are relevant: since both cones and convex
sets that contain the origin are star-shaped, both positively homogeneous
risk measures, such as Value-at-Risk, and convex risk measures, such as
Expected Shortfall, are star-shaped. Star-shaped sets and functions are well studied in mathematics; see, e.g., Rubinov and Yagubov \cite{RY86}.

It is safe to say that star-shaped risk measures encompass virtually all
 monetary (i.e., monotone and translation invariant) risk measures used in the literature and in the financial
practice. Perhaps more importantly, star-shaped risk measures, which are
in general not convex,   emerge in a number of relevant situations in risk management. 
In the next Section \ref{sec:ECB}, we
present several important examples illustrating the natural  appearance of these risk measures when computing margin charges in the presence of liquidity risk and competitive allocation, 
when different capital requirements are robustly aggregated,
when risk mitigation is possible for a decision maker, and when a non-concave utility function is involved in risk assessment.

The rest of the paper is dedicated to a comprehensive study of star-shaped
risk measures. Our main contributions are briefly summarized below.

In a style similar to the axiomatic foundation of risk measures due to
Artzner et al.~\cite{ADEH99} and F\"{o}llmer and Schied \cite{FS16}, the
main theory of star-shaped risk measures is developed in Sections \ref%
{sec:def} (basic definitions), \ref{sec:properties} (aggregative
operations), and \ref{sec:repr} (representation theorems). On the
mathematical side, we obtain an insightful minimax representation of
star-shaped risk measures as minima of convex risk measures (Theorem \ref%
{thm:conv}). \emph{This fact extends to any star-shaped risk
measure the elegant representation of Value-at-Risk as the minimum of all
convex risk measures that dominate it}, and actually characterizes
star-shaped risk measures through this envelope property. We also show that
the class of star-shaped risk measures is closed under the fundamental
aggregative operations of taking infima, suprema, mixtures, and
inf-convolutions (Theorem \ref{th:star_transf}), and that the corresponding
minimax representations admit very tractable formulas (Theorem \ref%
{th:rep-opre}). Section \ref{sec:optimization} studies the optimization
properties of star-shaped risk measures, while law-invariant star-shaped
risk measures are analyzed in Section \ref{sec:law}. All proofs are
relegated to the Appendix.

With this, we provide the general framework for star-shaped risk measures,
which, for the first time, unifies many risk measurement approaches that up to
the present day were considered irreconcilable: convex risk measures,
Value-at-Risk, and robust aggregation of risk measures; moreover, we connect them
to competitive risk sharing and to risk mitigation. Given the considerable
flexibility of this framework, many future research directions are possible,
and some of them are described in the concluding Section \ref%
{sec:conclusions}.
In this paper, we use the convention that a positive value of the realized random risk $X$ means a loss (i.e., a sign change from Artzner et al.~\cite{ADEH99}).

\section{Four motivating examples\label{sec:ECB}}

\subsection{Margin requirements with liquidity risk and multiple CCPs}
\label{sec:21}

In order to stabilize markets after the 2007-2008 financial crisis,
Dodd-Frank reform act in US and EMIR regulation in Europe impose that in
many derivative markets transactions are cleared by specialized
intermediaries, the so-called Central Counterparties (CCP herafter). In
order to reduce their exposure to insolvency risk, CCPs set a capital charge,
the margin requirement, to each market participant. Glassermann, Moallemi,
and Yuan \cite{GMY16} discuss margin settlement and explain that, to capture
liquidity costs at default, margin requirements at the level of individual
CCP need to be superlinear in position size. In the language of risk
measures, this is equivalent to requiring that the capital charge is
star-shaped with respect to the position size (see Proposition \ref%
{pro:star_char}), i.e., the individual CCP will set a risk capital charge
using a star-shaped risk measure. 

%

As observed by \cite{GMY16}, usually
dealers have access simultaneously to multiple competing CCPs and may split
the execution of a single deal across different CCPs. 
Next, we illustrate that in this setting, the effective
margin charge is star-shaped, but it may be non-convex even if each risk measure used by CCPs is
convex. Suppose that there are several
CCPs that may clear the position $X$ of a dealer, and assume that these CCPs are
associated with convex (hence star-shaped) risk measures $\rho _{1},\dots ,\rho _{n}$. If all of
the CCPs are simultaneously accessible, as in  \cite[(1)]{GMY16},  the dealer
would need to solve%
\begin{equation*}
\min_{X_{1},\dots ,X_{n}\in \mathcal{X}}\left\{ \sum_{i=1}^{n}\rho
_{i}(X_{i})~\Big |~\sum_{i=1}^{n}X_{i}=X\right\} 
\end{equation*}%
which is known as the operation of inf-convolution, discussed in Section \ref%
{sec:properties}.  
In practice, it may be infeasible or operationally costly
to trade with all CCPs at once; the dealer may face constraints on the configuration of CCPs that
can be involved for a specific deal.
\footnote{%
Some possible explanations of this restriction, quoting \cite[p.~1147]{GMY16}: \textquotedblleft Both parties to a swap need
to agree on where the swap will be cleared, and where their optimal
allocations may differ. To clear at a given CCP, both parties need to be
members of the CCP or trade through members of the CCP. [...] Clearing
members clear trades for clients as well as for their own accounts; this
limits their ability to subdivide positions. [...] Dealers may prefer one
CCP over another for reasons unrelated to margin requirements [...]" 
} Let $\mathcal{A}$ be the set of non-empty subsets of $\{1,\dots ,n\}$
representing compositions of CCPs the dealer can clear her position with.
This leads to the dealer's minimization problem 
\begin{equation*}
\min_{A\in \mathcal{A}}\min_{\substack{ X_{i}\in \mathcal{X} \\ 
		\text{s.t.~}i\in A}}\left\{ \sum_{i\in A}\rho _{i}(X_{i})~\Big |~\sum_{i\in
	A}X_{i}=X\right\} ,
\end{equation*}%
and its minimum value, denoted by $\rho (X)$, is the effective margin charge
for the position $X$. We can check that $\rho $ is a star-shaped risk measure
but not convex in general, because the minimum and the inf-convolution of
convex risk measures are star-shaped (see Theorem \ref{th:star_transf}) but
not necessarily convex due to the minimum operation over $\mathcal{A}$. %
In particular, if $\mathcal{A}$ contains only all sets with one element,
then $\rho $ is the minimum of $\rho _{1},\dots ,\rho _{n}$.

\subsection{Aggregation of convex risk measures and the ECB asset
purchase program}\label{sec:22}

Assume that a supervising agency consists of a board $I$ of experts, and
assume that each expert $i\in I$, based on her information and uncertainty attitudes,
proposes a convex risk measure $\rho _{i}$. Now the agency has to aggregate
these opinions, and of course any weighted average 
\begin{equation*}
\rho _{\mu }\left( X\right) =\sum_{i\in I}\mu _{i}\rho _{i}(X)
\end{equation*}%
is again a convex risk measure, where by a weighted average we mean that the $\mu _{i}$s are non-negative and
adding up to $1$. At the same time, it is well-known that averaging is
sensitive to outliers, and taking for instance a median is a popular robust
alternative.

Simple calculations show that a median, any order statistics, and any $L$%
-estimator of convex risk measures is a star-shaped risk measure, which is
typically not convex.\medskip 

To make the example applied to an economic context,
we consider the capital charge design in the European Central Bank (ECB) asset purchase program, 
an unconventional monetary policy with which the ECB
fosters the purchase of assets (issued both by private and public
institutions) to achieve the target inflation rate.

Let $I$ be the set of all national central
banks of the Euro-zone that participate in the   program. 
The program is designed according to a skin-in-the-game mechanism:
  a fraction of the losses deriving from a default of
the issuer of a purchased bond falls under a full European risk-sharing
regime, and the relevant    national central banks bear the remaining losses. 
 
How should the ECB define the capital charge required to face the resulting risk?
Assume that each national regulator  $i\in I$ adopts
a convex risk measure $\rho _{i}$ to assess the potential losses
arising from a portfolio $X$ of purchased assets. We consider the
selection a proper capital charge by the ECB for each purchase. 
In the extreme case of full loss sharing, the  precautionary amount
\begin{equation*}
\rho _{\vee }(X)=\max_{i\in I}\rho _{i} ( X ),
\end{equation*}%
 would be considered as a safe capital charge by all the involved national banks. In the opposite extreme case of no loss sharing, the potential loss
deriving from a default loss is bore in full by the national
central bank $j$ that will manage the execution of the purchase. In this
case, the ECB  acts like a ``dealer" in Section \ref{sec:21} by delegating
  risk management of the bond to the nation with minimal capital charge 
\begin{equation*}
\rho _{\wedge }(X)=\min_{i\in I}\rho _{i} ( X ).
\end{equation*}

Arguably, both rules  above are too extreme: the first immobilizes a great amount
of resources, and the second is quite fragile and does not exploit risk sharing. As a natural compromise, one may use
the convex combination 
\begin{equation*}
\rho \left(   X\right) = \mu \max_{i\in I}\rho
_{i}\left(   X\right) +(1-\mu )\min_{i\in I}\rho _{i}\left(   X\right) 
\end{equation*}%
as the effective capital charge, where the weight $\mu \in \left[ 0,1\right] $ has a clear interpretation as
an index of caution of the ECB. Of course, the resulting risk measure  $\rho$ is
star-shaped and not convex in general.

\subsection{Risk mitigation}
\label{sec:23}

From a decision theoretic perspective, a convex risk measure $\rho $ can be
seen as the certain equivalent loss of an ambiguity averse (and risk
neutral, as a financial institution is typically assumed to be) decision
maker who satisfies the rationality axioms of Maccheroni, Marinacci, and
Rustichini \cite{MMR06}. Since such a decision maker (DM) favors prospects $%
X $ with lower certain equivalent losses, his preferences are represented by
the certainty equivalent%
\begin{equation*}
\nu \left( X\right) =-\rho \left( X\right) =\min_{P\in \mathcal{M}}\left\{ 
\mathbb{E}_{P}[u\left( X\right) ]+\alpha \left( P\right) \right\}
\end{equation*}%
where $u( x) =-x$ because of risk neutrality and $\alpha (
P) $ is an ambiguity index that penalizes probabilistic scenarios
according to their plausibility. 
Formally, $\alpha $ is a convex and weak*-lower semicontinuous function on
the set $\mathcal{M}$ of probabilistic scenarios (finitely additive probabilities) which is \emph{grounded},
that is, such that $\min_{P\in \mathcal{M}}\alpha \left( P\right) =0$. 
Implicit in the uncertainty aversion axiom of \cite{MMR06} (which
corresponds to convexity of $\rho $) is the idea that a DM facing prospect $%
X $ cannot affect the plausibility of probabilistic scenarios.

As defended in a series of infuential papers by Jaques Dr\`{e}ze and Edi
Karni (see, e.g., Dr\`{e}ze \cite{Dre90} and Karni \cite{Kar11}), there are some
relevant cases in which a DM, by adopting a risk-mitigation strategy $a$,
may indeed affect such plausibility. In these cases, depending on $a$, the
certain equivalent loss is given by 
\begin{equation*}
\rho \left( a,X\right) =\max_{P\in \mathcal{M}}\left\{ \mathbb{E}%
_{P}[X]-\alpha _{a}\left( P\right) \right\}
\end{equation*}%
and the DM facing $X$ will choose $a$ to minimize it, so that his certainty
equivalent becomes%
\begin{equation}
\nu \left( X\right) =-\min_{a\in A}\rho \left( a,X\right) =\max_{a\in
A}\min_{P\in \mathcal{M}}\left\{ \mathbb{E}_{P}[u\left( X\right) ]+\alpha
_{a}\left( P\right) \right\}  \label{eq:starrep}
\end{equation}%
where $A$ is the set of risk-mitigation strategies (commonly called actions). 
 For this interpretation to be accurate it is necessary that the function $%
\rho \left( a,\cdot \right) $ be a \emph{bona fide} certainty equivalent
loss for each $a$, and in particular $\rho \left( a,c\right) =\rho \left(
b,c\right) =c$ for all actions $a,b\in A$ and every certain loss $c\in 
\mathbb{R}$. 
In other words, the requirement $\min_{P\in \mathcal{M}}\alpha _{a}\left(
P\right) =0$ for all $a\in A$, is what allows to interpret the
\textquotedblleft penalty functions\textquotedblright\ $\left\{ \alpha
_{a}:a\in A\right\} $ as ambiguity indexes rather than generic costs.

As observed in the previous example, $\min_{a\in A}\rho \left( a,\cdot \right) $ is
a star-shaped risk measure, which is not necessarily convex. Thus the
preferences 
\begin{equation*}
X\succsim Y\iff \nu \left( X\right) \geq \nu \left( Y\right)
\end{equation*}%
are not necessarily uncertainty averse. Yet, these preferences satisfy%
\begin{equation}
X\succsim c\implies \beta X+\left( 1-\beta \right) c\succsim c
\label{eq:irra}
\end{equation}%
for every uncertain prospect $X\in \mathcal{X}$, every certain prospect $%
c\in \mathbb{R}$, and every weight $\beta \in \left( 0,1\right) $, thus they
are \emph{increasingly relative ambiguity averse }in the sense of Xue \cite%
{X2020} and Cerreia-Vioglio, Maccheroni, and Marinacci \cite{CMM20}.%
\footnote{%
Indeed, an implication of our main Theorem \ref{thm:conv} is that the
preferences obtained by replacing uncertainty aversion with increasing
relative ambiguity aversion, in the axiom set of \cite{MMR06}, are
characterized by representation (\ref{eq:starrep}). See also Chandrasekher
et al. \cite{CFIL20} for a preference representation which is similar, but
where the $\alpha _{a}$'s are not necessarily grounded, hence they lose the
interpretation of action-dependent ambiguity indexes, and increasing
relative ambiguity aversion is lost too.} This simple observation is
important, in that it clarifies the decision-theoretic appeal of star-shaped
risk measures. They describe institutional decision makers who, in the face
of uncertainty, will take any available measure to mitigate its adverse
effects, and who, as the total capital of the financial institution
increases, will decrease the fraction of it that is exposed to those effects
(see the seminal Arrow \cite{A1971}).

\subsubsection*{Dr\`{e}ze meets Hansen and Sargent}

The first axiomatic treatment of risk-mitigation dates back to Dr\`{e}ze
(1987). In his work, for every uncertain prospect $X\in \mathcal{X}$,%
\begin{equation}
\nu \left( X\right) =\max_{a\in A}\mathbb{E}_{Q_{a}}[u\left( X\right) ]
\label{eq:dre}
\end{equation}%
this means that Dr\`{e}ze's decision makers are confident that by choosing $%
a\in A$ they can induce $Q_{a}$ without error. Hansen and Sargent \cite{HS01}
would say \textquotedblleft without fear of
misspecification.\textquotedblright\ They also show that misspecification
concerns on part of DMs can be addressed through robustification. Formally,
this is achieved by replacing $\mathbb{E}_{Q_{a}}[u\left( X\right) ]$ in (%
\ref{eq:dre}) with 
\begin{equation*}
-\lambda \log \mathbb{E}_{Q_{a}}\left[ \exp \left( -\frac{u\left( X\right) }{%
\lambda }\right) \right] =\min_{P\in \mathcal{M}}\left\{ \mathbb{E}%
_{P}[u\left( X\right) ]+\lambda \mathcal{R}\left( P\mid Q_{a}\right) \right\}
\end{equation*}%
where $\lambda >0$ captures the level of trust in the relation between $a$
and $Q_{a}$, and $\mathcal{R}$ is the relative entropy. This adjustment
leads to 
\begin{equation*}
\nu _{\lambda }\left( X\right) =\max_{a\in A}\min_{P\in \mathcal{M}}\left\{ 
\mathbb{E}_{P}[u\left( X\right) ]+\lambda \mathcal{R}\left( P\mid
Q_{a}\right) \right\}
\end{equation*}%
which for $\lambda =+\infty $ corresponds to Dr\`{e}ze's initial proposal,
while lower values of $\lambda $ correspond to lower confidence on part of
DMs in their ability to affect probabilistic scenarios. Clearly, this is a
special case of (\ref{eq:starrep}) with $\alpha _{a}\left( \cdot \right)
=\lambda \mathcal{R}\left( \cdot \mid Q_{a}\right) $.

\subsection{Non-concave utilities}\label{ex:fried}

F\"{o}llmer and Schied \cite{FS02,FS16} proposed the class of
utility-based shortfall risk measures 
\begin{equation}
\rho _{u}(X)=\inf \{m\in \mathbb{R}\mid \mathbb{E}_{P}[u(m-X)]\geq
u(0)\}\qquad X\in \mathcal{X}  \label{eq:shortfall}
\end{equation}%
where $P$ is a given probability measure and $u$ is an increasing and
non-constant utility function on $\mathbb{R}$ such that (w.l.o.g.) $u(0)=0$.
The interpretation of $\rho _{u}$ in \eqref{eq:shortfall} is that the
acceptable risk positions are the ones which have non-negative reservation
price. Note that \eqref{eq:shortfall} always defines a risk measure (in the formal sense of Definition \ref{def:convex} below) which is convex
if and only if $u$ is concave (see, e.g., \cite[Section 4.9]{FS16}).

Concavity of $u$ corresponds to a strong form of risk aversion. Utility
functions with local convexities have been studied and normatively justified
since Friedman and Savage \cite{FS48}. Motivated by this, Landsberger and
Meilijson \cite{LM90} and \cite{LM90b} consider the broader class of utility
functions such that 
\begin{equation}
\lambda \mapsto \frac{u(\lambda )}{\lambda }%
\mbox{~is decreasing  on $
(0,\infty)$ and $ (-\infty,0)$}  \label{eq:shortfall2}
\end{equation}%
which allow for convex kinks (see also M\"{u}ller \cite{M98} and M\"{u}ller et al.\ \cite{MSTW17}),
and they show that these functions capture \textquotedblleft aversion to
fatter\ profit/loss tails.\textquotedblright\ Note that \eqref{eq:shortfall2}
is equivalent to 
\begin{equation}
\lambda \mapsto \frac{u(\lambda x)}{\lambda }%
\mbox{~is decreasing  on $ 
(0,\infty)$   for all $x\in \mathbb R$}.  \label{eq:shortfall3}
\end{equation}

If $u$ satisfies \eqref{eq:shortfall2}, then the corresponding utility-based
shortfall risk measure $\rho _{u}$ in \eqref{eq:shortfall} is star-shaped.
This claim can be verified by, for $\lambda >1$, 
\begin{align*}
\rho _{u}(\lambda X)& =\inf \{\lambda m\in \mathbb{R}\mid \mathbb{E}%
_{P}[u(\lambda m-\lambda X)]\geq 0\} \\
& =\lambda \inf \left\{ m\in \mathbb{R}\mid \mathbb{E}_{P}\left[ \frac{%
u(\lambda m-\lambda X)}{\lambda }\right] \geq 0\right\}  \\
& \geq \lambda \inf \{m\in \mathbb{R}\mid \mathbb{E}_{P}[u(m-X)]\geq
0\}=\lambda \rho_u (X)
\end{align*}%
where the inequality is due to \eqref{eq:shortfall3}. Indeed, it is not
difficult to verify the stronger statement that \eqref{eq:shortfall} defines
a star-shaped risk measure $\rho _{u}$ if and only if $u$ satisfies \eqref{eq:shortfall2}.
Therefore, star-shaped risk measures arise naturally if non-concave
utility functions are involved in utility-based shortfall risk measurement 
\emph{\`{a} la} F\"{o}llmer and Schied.

%
%

\section{Star-shaped risk measures: basic definitions} 

\label{sec:def}

The points of departure of our analysis are the standard definitions that
shaped the theory of risk measurement as originally introduced in the
landmark papers by Artzner et al. \cite{ADEH99}, Delbaen \cite{D02}, F\"{o}%
llmer and Schied \cite{FS02}, and Frittelli and Rosazza Gianin \cite{FR02}.
As mentioned, we use the convention of {McNeil, Frey, and Embrechts }\cite%
{MFE15} that random variables represent future \emph{losses} of positions
held by financial institutions over some time horizon $T$. 
In particular, for a position with loss given by $X$, a negative value of $%
X(\omega )$ corresponds to a gain (if $\omega $ occurs).  The time horizon $%
T $ is left unspecified and, in order to simplify the presentation, we set
interest rates equal to zero so that there is no discounting.

Specifically, the possible losses of financial positions are represented by
a linear space $\mathcal{X}$ of bounded random variables containing all
constants. We do not assume that a probability measure is \emph{a priori}
given on the underlying measurable space $\Omega $ of states of the
environment. The space $\mathcal{X}$ is endowed with the pointwise order, so
that $X\geqq Y$ if, and only if, $X(\omega )\geq Y(\omega )$ for all states $%
\omega $.

\begin{definition}
\label{def:convex} 
A \emph{risk measure} is a function $\rho :\mathcal{X}%
\rightarrow \mathbb{R}$ that satisfies  
\begin{itemize}
\item[1.] \emph{Monotonicity:} If $X\geqq Y$, then $\rho (X)\geq \rho (Y)$%
;  

\item[2.] \emph{Translation invariance:} $\rho (X-m)=\rho (X)-m$ for all $%
X\in \mathcal{X}$ and all $m\in \mathbb{R}$;  

\item[3.] \emph{Normalization:} $\rho (0)=0$.  
 \end{itemize} 
A risk measure may satisfy the following further properties
\begin{itemize}
\item[4.] \emph{Star-shapedness:} $\rho (\lambda X)\geq \lambda \rho (X) $
for all $X\in \mathcal{X}$ and all $\lambda >1$; 
\item[5.] \emph{Convexity:} $\rho (\lambda X +(1-\lambda) Y) \le \lambda \rho (X) +(1-\lambda)\rho(Y)$ for all $X,Y\in \mathcal{X}$ 
  and all $\lambda \in (0,1)$;  
\item[6.] \emph{Positive homogeneity:} $\rho (\lambda X)=\lambda \rho (X)$
for all $X\in \mathcal{X}$ and all $\lambda >0$;    
\item[7.] \emph{Subadditivity:} $\rho (X+Y)\leq \rho (X)+\rho \left(
Y\right) $ for all $X,Y\in \mathcal{X}$.
\end{itemize}
 A risk measure $
\rho $  is \emph{coherent} if it satisfies  positive homogeneity and subadditivity.

\end{definition}


Introduced by \cite{FS02} and \cite{FR02}, convex risk measures are a very
popular class of risk measures that capture preference for diversification; 
see also Cerreia-Vioglio et al. \cite{CMMM11} for quasi-convex risk measures.
As anticipated, convex risk
measures are known to be star-shaped. In fact, for all $X\in \mathcal{X}$
and all $\lambda >1$, 
\begin{equation*}
\rho (X)=\rho \left( \frac{1}{\lambda }\left( \lambda X\right) +\left( 1-%
\frac{1}{\lambda }\right) 0\right) \leq \frac{1}{\lambda }\rho \left(
\lambda X\right) +\left( 1-\frac{1}{\lambda }\right) \rho (0)=\frac{1}{%
\lambda }\rho \left( \lambda X\right) .
\end{equation*}%
However, the converse is generally not true, because all positively
homogeneous risk measures are star-shaped too. For instance, Value-at-Risk
(VaR) is positively homogeneous without being convex. 
Specifically, given a probability measure $Q$,
VaR at level $\beta \in    ( 0,1 ] $ is defined by 
\begin{equation*}
\mathrm{VaR}_{\beta }^{Q}(X)=\inf \{x\in \mathbb{R}:Q(X>x)\le 1-\beta \}
\end{equation*}%
and it is the minimum capital reserve that brings default probability below $%
1-\beta $. Non-convexity of VaR led to the introduction of Expected
Shortfall (ES)
\begin{equation*}
\mathrm{ES}_{\beta }^{Q}(X)=\frac{1}{1-\beta }\int_{\beta }^{1}\mathrm{VaR}%
_{t}^{Q}(X)\mathrm{d}t
\end{equation*}%
which of course is star-shaped too. 
ES is the standard risk measure in Basel IV and has been recently axiomized by Wang and Zitikis \cite%
{WZ21}. 

A main concern regarding both $\mathrm{VaR}_{\beta }^{Q}(X)$ and $\mathrm{ES}%
_{\beta }^{Q}(X)$ is the fact that their value depends crucially on $Q$,
thus the possible misspecification of $Q$ makes these measures very fragile.
This, in turn, led recent research on risk measures to investigate
robustifications like the ones described in the next two examples. Let us
anticipate that these robustifications generate monetary risk measures which
are always star-shaped and typically non-convex.

\begin{example}[Scenario-based risk measures]
\label{ex:1}Wang and Ziegel \cite{WZ18} consider a collection $\mathcal{Q}$
of probability measures and define 
\begin{equation*}
\mathrm{MaxVaR}_{\beta }^{\mathcal{Q}}(X)=\sup_{Q\in \mathcal{Q}}\mathrm{VaR}%
_{\beta }^{Q}(X).
\end{equation*}%
This risk measure has a natural interpretation in terms of robustness, it is
widely used in applications (see, e.g., Natarajan et al. \cite{NPS08}), and like VaR it is generally not
convex, but being positively homogeneous it is star-shaped.

Alternatively, for the finite $\mathcal{Q}$'s that typically appear in
applications, one might consider the less extreme robustification 
\begin{equation*}
\mathrm{MedVaR}_{\beta }^{\mathcal{Q}}(X)=\mathrm{Median}\{\mathrm{VaR}%
_{\beta }^{Q}(X)\mid Q\in \mathcal{Q}\}.
\end{equation*}%
Like $\mathrm{MaxVaR}_{\beta }^{\mathcal{Q}}$ also $\mathrm{MedVaR}_{\beta
}^{\mathcal{Q}}(X)$ is positively homogeneous and generally not convex.

When VaR is replaced by ES, it is easy to show that $\mathrm{%
MaxES}_{\beta }^{\mathcal{Q}}(X)$ is a coherent risk measure (like ES) while 
$\mathrm{MedES}_{\beta }^{\mathcal{Q}}(X)$ is star-shaped without being
coherent or even convex.
\end{example}

The next example goes one step further by yielding a non-convex and
non-positively homogeneous, yet very much tractable risk measure.

\begin{example}[Benchmark loss VaR]
\label{ex:2} The benchmark loss VaR in Bignozzi et al. \cite{BBM20} is
defined as 
\begin{equation*}
\mathrm{LVaR}_{\alpha }^{Q}(X)=\sup_{t\geq 0}\{\mathrm{VaR}_{\alpha
(t)}^{Q}(X)-t\}
\end{equation*}%
where $\alpha :[0,\infty )\rightarrow (0,1]$ is an increasing and
right-continuous function. It is easy to check that $\mathrm{LVaR}_{\alpha
}^{Q}$ is neither positively homogeneous nor convex, and yet it is a
star-shaped risk measure.
\end{example}

Next we formalize some properties of star-shaped risk measures that we
informally anticipated in the introduction. We begin with a simple, but
conceptually important characterization in terms of risk-to-exposure ratios.

\begin{proposition}
\label{pro:star_char}For a risk measure $\rho :\mathcal{X}\rightarrow 
\mathbb{R}$, the following are equivalent:

\begin{enumerate}
\item[(i)] $\rho $ is star-shaped;

\item[(ii)] $\rho (\alpha X)\leq \alpha \rho (X)$ for all $X\in \mathcal{X}$
and all $\alpha \in \left( 0,1\right) $;

\item[(iii)] for each $X\in \mathcal{X}$, the risk-to-exposure ratio $%
r_{X}:\beta \mapsto {\rho (\beta X)}/{\beta }$ is an increasing function of $%
\beta $ on $(0,\infty )$.
\end{enumerate}
\end{proposition}

These properties of star-shaped functions explain why they have also been
called \textquotedblleft increasing along rays\textquotedblright\ or
\textquotedblleft radiant\textquotedblright\ in the context of abstract
convex analysis, where this property was first discovered; 
see, e.g., Zaffaroni \cite{Z04} and Penot \cite{P05}. 

An acceptance set is naturally associated with each risk measure $\rho $. It
is the set of positions that do not require any additional capital.
Formally, it coincides with the lower level set of $\rho $ at $0$, that is, 
\begin{equation*}
\mathcal{A}_{\rho }=\left\{ X\in \mathcal{X}\mid \rho \left( X\right) \leq
0\right\} .
\end{equation*}%
As well known, $\mathcal{A}_{\rho }$ completely determines $\rho $; in fact,
the translation invariance property guarantees that%
\begin{equation}
\rho \left( X\right) =\min \left\{ m\in \mathbb{R}\mid X-m\in \mathcal{A}%
_{\rho }\right\} \qquad X\in \mathcal{X}.  \label{eq:tri}
\end{equation}%
In other words, the risk of $X$ is measured as the minimum amount by which
the loss it represents must be uniformly reduced to make the adjusted
position acceptable.

In general, a subset $\mathcal{A}$ of $\mathcal{X}$ such that:%
\begin{gather*}
\sup \left\{ m\in \mathbb{R}\mid m\in \mathcal{A}\right\} =0\text{ and} \\
X\in \mathcal{A},\ Y\in \mathcal{X},\ Y\leqq X\implies Y\in \mathcal{A}
\end{gather*}%
is called an \emph{acceptance set}, and it generates a risk measure 
\begin{equation*}
\rho _{\mathcal{A}}\left( X\right) =\inf \left\{ m\in \mathbb{R}\mid X-m\in 
\mathcal{A}\right\} \qquad X\in \mathcal{X}
\end{equation*}%
which is convex (resp., positively homogeneous) if $\mathcal{A}$ is convex
(resp., a cone); see F\"{o}llmer and Schied \cite{FS16} for details. For obvious reasons, it
is convenient to call \emph{coherent} an acceptance set which is a convex
cone.

The fact that similar relations hold between star shapedness of a risk
measure and star shapedness of its acceptance sets, shows that a risk
measure is star-shaped if and only if it is based on the following
principle: deleveraging an acceptable position cannot make
it unacceptable.  
Equivalently, increasing the exposure to an unacceptable position cannot make it acceptable. 

Below, recall that a subset $S$ of a vector space is \emph{star-shaped} if, and
only if, $\lambda s\in S$ for all $\lambda \in \left[ 0,1\right] $ and all $%
s\in S$.

\begin{proposition}
\label{pro:star_acc}For a risk measure $\rho :\mathcal{X}\rightarrow \mathbb{%
R}$, the following are equivalent:

\begin{enumerate}
\item[(i)] $\rho $ is star-shaped;

\item[(ii)] the set $\mathcal{A}_{\rho }$ is star-shaped in $\mathcal{X}$;

\item[(iii)] there exists a star-shaped acceptance set $\mathcal{A}$ such
that $\rho =\rho _{\mathcal{A}}$.
\end{enumerate}
\end{proposition}

Finally, coherent risk measures coincide with subadditive and star-shaped
risk measures; thus here positive homogeneity
can be replaced by the weaker
property of star shapedness. This fact sheds some light on the strengths and weaknesses
of the subadditivity assumption.

\begin{proposition}
\label{pro:homo_super}For a subadditive risk measure $\rho :\mathcal{X}%
\rightarrow \mathbb{R}$, the following are equivalent:

\begin{enumerate}
\item[(i)] $\rho $ is star-shaped;

\item[(ii)] $\rho $ is positively homogeneous (thus coherent);

\item[(iii)] $\rho $ is convex.
\end{enumerate}
\end{proposition}

\section{Aggregation operations}

\label{sec:properties}

As explained in Section \ref{sec:ECB}, star-shaped risk measures naturally
emerge when a collection of risk measures $\left\{ \rho _{i}\right\} _{i\in
I}$ have to be aggregated in single risk measure $\rho $. To analyze these
situations, we introduce some operations that cover a broad range of
aggregators.

\begin{itemize}
\item For each probability $\mu $ on the parts of $I$, the average $\rho
_{\mu }$, is defined as 
\begin{equation}
\rho _{\mu }(X)=\int_{I}\rho _{i}(X)\,\mathrm{d}\mu \left( i\right) \qquad
X\in \mathcal{X}.  \label{eq:ave}
\end{equation}%
In particular, if $I$ is finite or $\mu $ is supported on a finite subset of 
$I$, then $\rho _{\mu }$ is a convex combination of $\left\{ \rho
_{i}\right\} _{i\in I}$, i.e., $\rho _{\mu }=\sum_{i\in I}c_{i}\rho _{i}$
where $c_{i}=\mu (\{i\})$, $i\in I$.

\item The supremum $\rho _{\vee }$, is defined as 
\begin{equation*}
\rho _{\vee }(X)=\sup_{i\in I}\rho _{i}(X)\qquad X\in \mathcal{X}.
\end{equation*}

\item The infimum $\rho _{\wedge }$, is defined as 
\begin{equation*}
\rho _{\wedge }(X)=\inf_{i\in I}\rho _{i}(X)\qquad X\in \mathcal{X}.
\end{equation*}

\item The inf-convolution $\rho _{\diamond }$, is defined as 
\begin{equation}
\rho _{\diamond }(X)=\inf \left\{ \sum_{i\in I}\rho _{i}\left( Y_{i}\right)
\ \left\vert \ Y_{i}\in \mathcal{X}\text{ for all }i\in I\text{ and\ }%
\sum_{i\in I}Y_{i}=X\right. \right\} \qquad X\in \mathcal{X}
\label{eq:splitting}
\end{equation}%
provided $I=\left\{ 1,2,...,n\right\} $ is finite and%
\begin{equation}
\sum_{i\in I}\rho _{i}\left( Z_{i}\right) \geq 0  \label{eq:split}
\end{equation}%
for all $Z_{1},Z_{2},...,Z_{n}\in \mathcal{X}$ such that $\sum_{i\in
I}Z_{i}=0$.\footnote{%
The normality condition (\ref{eq:split}) is required because also in the
case of two convex risk measures their inf-convolution might fail to be a
risk measure (see Barrieu and El Karoui \cite{BE05}).\ This condition is
obviously satisfied if there exists a linear functional which is dominated
by all risk measures in the collection.}
\end{itemize}

The following theorem shows that the class of star-shaped risk measures is
closed under these operations.

\begin{theorem}
\label{th:star_transf}For a collection of star-shaped risk measures, their
average, supremum, infimum, and inf-convolution (when defined) are
star-shaped risk measures.
\end{theorem}

Theorem \ref{th:star_transf} implies that star-shaped risk measures form a
complete lattice, i.e., a set that is closed with respect to the pointwise supremum
and infimum operations. Moreover, the inf-convolution risk measure is
important because it can be considered as the one used by a representative
agent in a risk-sharing or order-splitting problem (see, e.g., Embrechts et
al. \cite{ELW18} and the references therein), and star shapedness is
preserved in such situations.

In the proof of Theorem \ref{th:star_transf} we show that (\ref{eq:ave})
defines a star-shaped risk measure even if $\mu $ is a capacity on $I$ and
the integral is in the sense of Choquet
(a capacity on $I$ is a set function on a $\sigma$-field of $I$ such that $0=\mu \left(
\varnothing \right) \leq \mu \left( J\right) \leq \mu \left( K\right) \leq
\mu \left( I\right) =1$ for all $J\subseteq K\subseteq I$).  This is not done
for mathematical elegance, but because: first, the supremum case $\rho
_{\vee }$ and the infimum case $\rho _{\wedge }$ are special cases of
Choquet averages; second, when $I$ is finite, any order statistic (such as
the median) of $\left\{ \rho _{i}(X)\right\} _{i\in I}$ has form (\ref%
{eq:ave}) for a suitable capacity $\mu $ (see Murofushi and Sugeno \cite%
{MS93}).

\begin{remark}
Theorem \ref{th:star_transf} and what we have just observed about Choquet
averages show that the class of star-shaped risk measures is closed under
many commonly used economic aggregation mechanisms that disrupt convexity
(see also Cerreia-Vioglio, Corrao, and Lanzani \cite{CCL20}). Later we will
see in Section \ref{sec:law} that some subclasses of star-shaped risk
measures, in particular, law-invariant ones and $\mathcal{SSD}$-consistent
ones, are also closed under the same operations.
\end{remark}

Finally, the inf-convolution $\rho _{\diamond }$ is denoted by $\square
_{i\in I}\rho _{i}$ when it is necessary to make explicit the collection $%
\left\{ \rho _{i}\right\} _{i\in I}$ of risk measures that it aggregates.

\section{Representation of star-shaped risk measures}

\label{sec:repr}

Proposition 4.47 of F\"{o}llmer and Schied \cite{FS16}\ shows that VaR is
the minimum of the collection of all convex risk measures that dominate it.
Our main representation theorem shows that, not only this is true for all
star-shaped risk measures, but this property actually characterizes them.

\begin{theorem}
\label{thm:conv} For a risk measure $\rho :\mathcal{X}\rightarrow \mathbb{R}$%
, the following are equivalent:

\begin{enumerate}
\item[(i)] $\rho $ is star-shaped (resp.~positively homogeneous);

\item[(ii)] there exists a collection $\Gamma $ of convex (resp.~coherent)
risk measures such that%
\begin{equation}
\rho (X)=\min_{\gamma \in \Gamma }\gamma (X)\qquad X\in \mathcal{X};
\label{eq:attain}
\end{equation}

\item[(iii)] there exists a family $\left\{ \mathcal{A}_{\beta }\right\}
_{\beta \in B}$ of convex (resp.~coherent) acceptance sets such that%
\begin{equation*}
\rho (X)=\min \left\{ m\in \mathbb{R}\mid X-m\in \mathcal{A}_{\beta }\text{
for some }\beta \in B\right\} \qquad X\in \mathcal{X}.
\end{equation*}
\end{enumerate}
Moreover, in (ii) $\Gamma $ can be chosen as the collection of all convex (resp.~coherent) risk measures dominating $\rho $ and in (iii) $\left\{ \mathcal{A}%
_{\beta }\right\} _{\beta \in B}$ can be chosen as the family of their
acceptance sets.
\end{theorem}

The intuition behind Theorem \ref{thm:conv}, in particular the 
implication (i)$\Rightarrow$(ii), is best explained with
Figure \ref{fig:1} in the simple case of  $|\Omega|=2$ (without positive homogeneity).
The main idea is to write the acceptance set $\mathcal A_\rho$ of a star-shaped risk measure $\rho$ as the union of convex sets $\mathcal A_Y$ for $Y\in \mathcal X$, where $\mathcal A_Y$ contains all random variables dominated by $\alpha (Y-\rho(Y))$ for some $\alpha\in [0,1]$. 
As such, $\rho$ can be written as the minimum of convex risk measures $\rho_Y$, each associated with the acceptance set $\mathcal A_Y$.
 
\def\chiright{30.14}
\def\xzero{29.07}
\def\grauliber{19}

\pgfplotsset{
  myplot/.style={xtick=\empty, ytick=\empty, width=12.5cm, height=4.5cm}
}

\begin{figure}[!htb]  
\centering
\begin{tikzpicture}
[scale=.28,description/.style=auto]  
\draw [top color=gray!50, bottom color=gray!50, white] (-3,2)--(0,0)--(0,-18)--(-3,-19)--(-3,2); 
\fill [top color=gray, bottom color=gray!40] (-21.5,2)--(-21,-7)--(-17,-14)--(-11,-17)--(-3,-19)--(-3,2)--(-21.5,2);
\draw[thin] (0,0)--(-3,2)--(-3,-19);
\draw[thin] (-3,2)--(-22,2);
\draw[-latex] (0,-19) -- (0,11);
\draw[-latex] (-25,0) -- (16,0);

\node   at (1.2,10){{\small $\omega_2$}};

\node   at (15,-1){{\small $\omega_1$}};

 \fill [white,opacity=0.4,postaction={pattern=north east lines}] (-22,6.2)--(-15,5)--(-9,3)--(-6,3)--(-4,2.1)--(-3,2)--(-21.5,2)--(-21.5,6.2);
  \fill [white,opacity=0.4,postaction={pattern=north east lines}] (-3,2)--(-1,1.8)--(0,0)--(-3,2);
 \fill [white,opacity=0.4,postaction={pattern=north east lines}] (0,0)--(2,-1.8)--(2,-4)--(3,-6)--(3,-9)--(4,-16)--(0,-18)--(0,0);
  
 \draw[very thick] (-21.5,6.2)--(-15,5)--(-9,3)--(-6,3)--(-4,2.1)--(-3,2)--(-1,1.8)--(0,0)--(2,-1.8)--(2,-4)--(3,-6)--(3,-9)--(4,-16);
\draw[very thick, dashed] (-3,2)--(2,7);
\node (a0) at (3.3,7.3){{\small ${Y}$}};
\node (a2) at (0,0){$\bullet$};
\node (a3) at (.8,.8){{\small ${0}$}};
\node (a4) at (-3,2){$\bullet$};
\node (a5) at (2.1,7){$\bullet$};
        \path[<-, draw,  thick, dashed] (2.25,-3.5)  to[out=30, in=90] (11,-8)      node[below] 
        {\small boundary of $\mathcal A_\rho$};   
 
    \path[<-, draw,  thick, dashed] (1.4,-10)
      to[out=270, in=180] (5.5,-18)
      node[right] {\small $\mathcal A_\rho\setminus \mathcal A_Y$}; 
          \path[<-, draw,  thick, dashed] (-10,-7)
      to[out=225, in=90] (-14,-17)
      node[below] {\small $\mathcal A_{Y}$}; 
          \path[<-, draw,  thick, dashed] (-3.35,2.35)
      to[out=135, in=245] (-5,8)
      node[above] {\small$Y-\rho(Y)$}; 
\end{tikzpicture}
\caption{The intuition behind Theorem \ref{thm:conv} in case $\Omega=\{\omega_1,\omega_2\}$, where $\mathcal A_{Y}$ is the set of all random variables dominated by   $\alpha(Y-\rho(Y))$ for some $\alpha\in[0,1]$.}
\label{fig:1}
\end{figure}
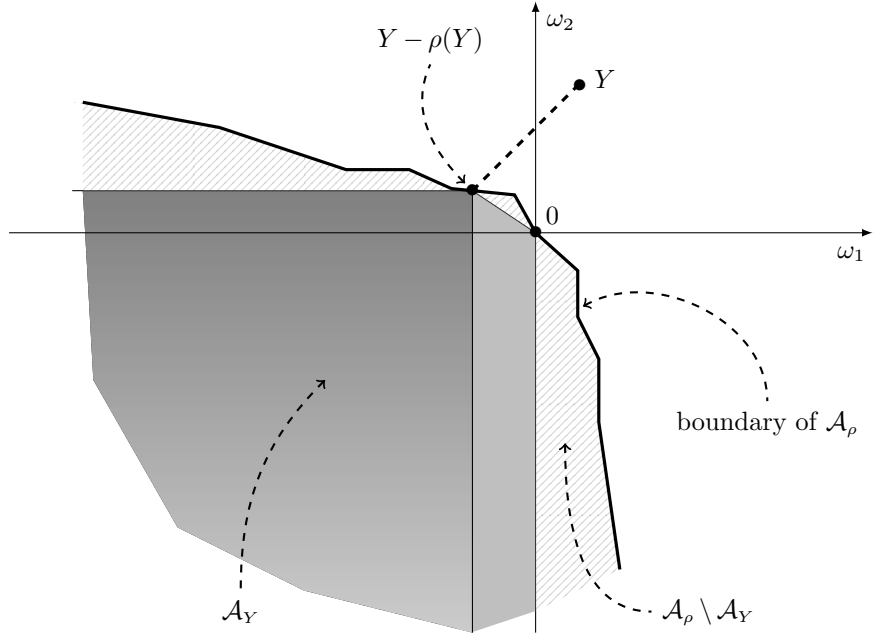

Theorem \ref{thm:conv} yields a tractable representation of star-shaped risk
measures, and its first use in applications dates back to Castagnoli et al. 
\cite{Cal15} (where it is stated without proof as Proposition 2). The special case of
coherent risk measures with $\mathcal{X}=\mathbb{R}^{n}$ also appears in the
independent Chandrasekher et al. \cite{CFIL20}.

As it happens for convex (and coherent) risk measures this envelope
representation is not unique, unless a suitable relaxation is considered.
For each set of convex (resp.~coherent) risk measures $\Gamma $ define its
relaxation $\tilde{\Gamma}$ by%
\begin{equation*}
\tilde{\Gamma}=\left\{ \tilde{\gamma}:\mathcal{X}\rightarrow \mathbb{R}\
\left\vert \ 
\begin{array}{l}
\tilde{\gamma}\text{ is a convex (resp.~coherent) risk measure and} \\ 
\text{for each }X\in \mathcal{X}\ \text{there is }\gamma \in \Gamma \text{
such that }\tilde{\gamma}\left( X\right) \geq \gamma \left( X\right)%
\end{array}%
\right. \right\} .
\end{equation*}

\begin{proposition}
\label{prop:uni}Let $\rho $ be a star-shaped risk measure and $\Gamma _{1}$
and $\Gamma _{2}$ be two sets of convex risk measures such that%
\begin{equation*}
\rho (X)=\min_{\gamma _{1}\in \Gamma _{1}}\gamma _{1}(X)=\min_{\gamma
_{2}\in \Gamma _{2}}\gamma _{2}(X)\qquad X\in \mathcal{X}
\end{equation*}%
then $\tilde{\Gamma}_{1}=\tilde{\Gamma}_{2}=\left\{ \gamma :\mathcal{X}%
\rightarrow \mathbb{R}\mid \gamma \text{ is a convex risk measure and }%
\gamma \geqq \rho \right\} $.

In particular, the set $\tilde{\Gamma}=\tilde{\Gamma}_{1}=\tilde{\Gamma}_{2}$
does not depend on the selected representation, but only on $\rho $.
\end{proposition}

The next result shows that the representation obtained in Theorem \ref%
{thm:conv}, is well behaved with respect to the aggregative operations
described in Section \ref{sec:properties}.

\begin{theorem}
\label{th:rep-opre}Let $\left\{ \rho _{i}\right\} _{i\in I}$ be a collection
of star-shaped risk measures and $\left\{ \Gamma _{i}\right\} _{i\in I}$ a
collection of sets of convex risk measures such that, for each $i\in I$,%
\begin{equation}
\rho _{i}\left( X\right) =\min_{\gamma _{i}\in \Gamma _{i}}\gamma _{i}\left(
X\right) \qquad X\in \mathcal{X}.  \label{eq:op0}
\end{equation}%
The average, supremum, infimum, and inf-convolution (when defined) of the
collection $\left\{ \rho _{i}\right\} _{i\in I}$ are given by 
\begin{align}
\int_{I}\rho _{i}(X)\,\mathrm{d}\mu \left( i\right) & =\min_{\gamma \in \int
\Gamma _{i}\,\mathrm{d}\mu \left( i\right) }\gamma (X)\qquad \text{where }%
\int \Gamma _{i}\,\mathrm{d}\mu \left( i\right) =\left\{ \int_{I}\gamma
_{i}\,\mathrm{d}\mu \left( i\right) \ \left\vert \ \gamma _{i}\in \Gamma _{i}%
\text{ for all }i\in I\right. \right\}  \label{eq:op1} \\
\sup_{i\in I}\rho _{i}(X)& =\min_{\gamma \in \cap _{i}\tilde{\Gamma}%
_{i}}\gamma (X)  \label{eq:op2} \\
\inf_{i\in I}\rho _{i}(X)& =\inf_{\gamma \in \cup _{i}\Gamma _{i}}\gamma (X)
\label{eq:op3} \\
\rho _{\diamond }(X)& =\inf_{\gamma \in \square _{i\in I}\Gamma _{i}}\gamma
(X)\qquad \text{where }\square _{i\in I}\Gamma _{i}=\left\{ \square _{i\in
I}\gamma _{i}\mid \gamma _{i}\in \Gamma _{i} \text{ for all }i\in I\right\} .  \label{eq:op4}
\end{align}%
for all $X\in \mathcal{X}$. 
Moreover, the infimum on the r.h.s.~of  \eqref{eq:op3}  is attained when $\inf_{i\in I}\rho
_{i}(X)$ is attained on $I$, and the infimum  on the r.h.s.~of  \eqref{eq:op4}  is attained
when%
\begin{equation*}
\rho _{\diamond }(X)=\min \left\{ \sum_{i\in I}\rho _{i}\left( Y_{i}\right)
\ \left\vert \ Y_{i}\in \mathcal{X}\text{ for all }i\in I\text{ and\ }%
\sum_{i\in I}Y_{i}=X\right. \right\}
\end{equation*}%
that is, if the infimum in the definition of $\rho _{\diamond }$ is attained.
\end{theorem}

\begin{remark}
\label{rem:ata}In formula (\ref{eq:op2}), it is in general necessary to use
the relaxations $\tilde{\Gamma}_{i}$ of the original sets $\Gamma _{i}$. A
counterexample to the use of the $\Gamma _{i}$'s can be found by choosing $%
\Gamma _{1}=\left\{ \gamma _{1}\right\} $ and $\Gamma _{2}=\left\{ \gamma
_{2}\right\} $ with $\gamma _{1}\neq \gamma _{2}$ so that $\Gamma _{1}\cap
\Gamma _{2}=\varnothing $.
\end{remark}

The counterparts of the latter two statements for positively homogeneous
risk measures are straightforward. Finally, assume that $\mathcal{X}=L^{\infty }\left( \Omega ,\mathcal{F}%
,P\right) $ where $P$ is any given probability measure. 
Let $\mathcal{P%
}$ be the set of all probability measures on $(\Omega ,\mathcal{F)%
}$ which are absolutely continuous with respect to $P$. 
Any convex risk measure $\gamma$ on $\mathcal{X}$ 
with the Fatou property 
can be written
as  (\cite[Theorem 4.33]{FS16})%
\begin{equation*}
\gamma (X)=\sup_{Q\in \mathcal{P}}\left\{ \mathbb{E}_{Q}[X]-\alpha _{\gamma
}\left( Q\right) \right\}, \qquad X\in \mathcal{X},
\end{equation*}%
where $\alpha _{\gamma }:\mathcal{P}\rightarrow \left[ 0,\infty \right] $ is
such that $\inf_{Q\in \mathcal{P}}\alpha _{\gamma }\left( Q\right) =0$.
If the Fatou property is not assumed, then the set $\mathcal P$ needs to be replaced by the set of all finitely additive probabilities. 
In the next result, we obtain a robust representation of star-shaped risk measures based on  Theorem \ref{thm:conv}, 
in which each convex risk measure $\gamma$ satisfies the Fatou property and thus can be represented via $\alpha _{\gamma }:\mathcal{P}\rightarrow \left[ 0,\infty \right] $.

\begin{proposition}
\label{coro:robust}A risk measure $\rho :\mathcal{X}\rightarrow \mathbb{R}$
is star-shaped if and only if there exists a collection $\left\{ \alpha
_{\gamma }\right\} _{\gamma \in \Gamma }$ of functions $\alpha _{\gamma }:%
\mathcal{P}\rightarrow \left[ 0,\infty \right] $, with $\inf_{Q\in \mathcal{P%
}}\alpha _{\gamma }\left( Q\right) =0$ for all $\gamma \in \Gamma $, such
that%
\begin{equation}
\label{eq:rep-FS}
\rho (X)=\min_{\gamma \in \Gamma }\sup_{Q\in \mathcal{P}}\left\{ \mathbb{E}%
_{Q}[X]-\alpha _{\gamma }\left( Q\right) \right\}, \qquad X\in \mathcal{X}.
\end{equation}%
\end{proposition}

\section{Optimization of star-shaped risk measures}

\label{sec:optimization} 

Theorem \ref{thm:conv} allows us to use convex optimization techniques to
optimize star-shaped risk measures. Let $\left\{ \rho _{i}\right\} _{i\in I}$
be a collection of star-shaped risk measures with representations%
\begin{equation*}
\rho _{i}(X)=\min_{\gamma \in \Gamma _{i}}\gamma (X)\qquad X\in \mathcal{X}
\end{equation*}%
where $\Gamma _{i}$ is a set of convex risk measures for all $i\in I$. Like
in Example \ref{ex:1}, $I=\mathcal{Q}$ may be a set of probability measures
on $\Omega $ and each $\rho _{Q}$ may be a risk measure which is
law-invariant under $Q$, say $\mathrm{VaR}_{\beta }^{Q}$.

Let $\boldsymbol{S}$ be a vector of risk factors, that is, random variables
on $\Omega $, $A$ be a set of available actions, and $\ell :A\times \mathbb{R%
}^{n}\rightarrow \mathbb{R}$ be a loss function. The natural interpretation
of $\ell (a,\boldsymbol{S})$ is as the random loss corresponding to action $%
a $ and risk factor $\boldsymbol{S}$. Consider the standard risk
minimization problem%
\begin{equation}
\mbox{to minimize }\quad
 \rho _{i}(\ell (a,\boldsymbol{S})) 
\quad\mbox{over $a\in A$}
 \label{eq:sopt1}
\end{equation}%
%
and its robust version%
\begin{equation}
\mbox{to minimize }\quad \sup_{i\in I}\rho _{i}(\ell (a,\boldsymbol{S}))\quad\mbox{over $a\in A$.}
\label{eq:sopt2}
\end{equation}%
Again note that the objective $\rho _{i}$ in (\ref{eq:sopt1}) and its robust
version $\rho _{\vee }=\sup_{i\in I}\rho _{i}$ in (\ref{eq:sopt2}) are
star-shaped, but not necessarily convex risk measures. Nevertheless, using
Theorems \ref{thm:conv} and \ref{th:rep-opre}, the two optimization problems
above can be converted into standard optimization problems for convex risk
measures.

\begin{proposition}
\label{prop:opt}If $\ell (a,\boldsymbol{S})\in \mathcal{X}$ for all $a\in A$%
, then%
\begin{equation}
\inf_{a\in A}\rho _{i}(\ell (a,\boldsymbol{S}))=\inf_{\gamma \in \Gamma
_{i}}\inf_{a\in A}\gamma (\ell (a,\boldsymbol{S}))  \label{eq:sopt3}
\end{equation}%
and 
\begin{equation}
\inf_{a\in A}\sup_{i\in I}\rho _{i}(\ell (a,\boldsymbol{S}))=\inf_{\gamma
\in \tilde{\Gamma}_{I}}\inf_{a\in A}\gamma (\ell (a,\boldsymbol{S})).
\label{eq:sopt4}
\end{equation}%
where $\tilde{\Gamma}_{I}=\bigcap_{i\in I}\tilde{\Gamma}_{i}$. Moreover,

\begin{enumerate}
\item[1.] $a^{\ast }$ is a minimizer of problem  \eqref{eq:sopt1} if, and only if,
there exists $\gamma ^{\ast }\in \Gamma _{i}$ such that $(a^{\ast },\gamma
^{\ast })$ minimizes $\gamma (\ell (a,\boldsymbol{S}))$ over $(a,\gamma )\in
A\times \Gamma _{i}$;

\item[2.] $a^{\ast }$ is a minimizer of problem \eqref{eq:sopt2}  if, and only if,
there exists $\gamma ^{\ast }\in \tilde{\Gamma}_{I}$ such that $(a^{\ast
},\gamma ^{\ast })$ minimizes $\gamma (\ell (a,\boldsymbol{S}))$ over $%
(a,\gamma )\in A\times \tilde{\Gamma}_{I}$.
\end{enumerate}
\end{proposition}

As a direct consequence of Proposition \ref{prop:opt}, the optimal financial
position for a star-shaped risk measure $\rho $ in a given subset is the
minimum among the optimal financial positions of the convex risk measures
which represent it in the sense of Theorem \ref{thm:conv}.

\begin{corollary}
\label{cor:optim}If $\rho :\mathcal{X}\rightarrow \mathbb{R}$ is a
star-shaped risk measure with representation \eqref{eq:attain} and $\mathcal{%
Y}\subseteq \mathcal{X}$, then%
\begin{equation*}
\inf_{X\in \mathcal{Y}}\rho (X)=\inf_{\gamma \in \Gamma }\inf_{X\in \mathcal{%
Y}}\gamma (X).
\end{equation*}
\end{corollary}

\begin{example}[Portfolio selection with risk measures]
An important problem in finance is portfolio selection with various
objectives and constraints. If the objective is to minimize a risk measure $%
\rho$, a classic problem can be formulated as 
\begin{equation}
\mbox{to minimize: } \rho(-Y_T) \mbox{~~~~over portfolios $Y$~~}%
\mbox{~~subject to~} Y_0\le x_0 \mbox{~and~$Y_T\in \mathcal Y$},
\label{eq:opt2}
\end{equation}
where $x_0$ is a constant representing the budget, $Y_T$ is the terminal
payoff at time $T$ of a dynamically traded portfolio process $Y$, $Y_0$ is
its initial price, and $\mathcal{Y}$ is a set of acceptable positions on the
terminal payoff; e.g., there may be a bound on the maximum loss from $Y_T$.

For the purpose of illustration, we assume a complete market in which the
risk-free interest rate is zero. In a complete market, it is well known that
problem \eqref{eq:opt2} can be solved in two steps (see e.g., F\"{o}llmer,
Schied, and Weber \cite{FSW09}). First, solve the static problem 
\begin{equation}
\mbox{to minimize: }\rho (-X)\mbox{~~~~over~}X\in \mathcal{X}%
\mbox{~~~~subject to~}\mathbb{E}^{Q}[X]\leq x_{0}%
\mbox{~and~$X\in \mathcal
Y$},  \label{eq:opt3}
\end{equation}%
where $Q$ is the unique martingale measure in the financial market, and $%
\mathcal{X}$ is the set of random variables representing time-$T$ payoffs.
Second, replicate a portfolio with $Y_{T}=X$ and $Y_{0}=\mathbb{E}^{Q}[X]$
using a standard method of martingale representation.

For various choices of $\mathcal{Y}$ and a convex risk measure $\rho $,
problem \eqref{eq:opt3} has been well studied and often admits explicit
solutions; see Chapter 8 of F\"{o}llmer and Schied \cite{FS16}. On the other
hand, analytical solutions to problem \eqref{eq:opt2} for non-convex risk
measures are rarely available as it involves non-convex optimization. Using
Corollary \ref{cor:optim}, we can solve \eqref{eq:opt3} for any star-shaped
risk measure $\rho $. First, we write $\rho =\min_{\gamma \in \Gamma }\gamma 
$ for some set $\Gamma $ of convex risk measures. Second, we solve for%
\begin{equation*}
X_{\gamma }^{\ast }\in \mathrm{argmin}\{\gamma (-X):X\in \mathcal{X}\cap 
\mathcal{Y}\text{\ and }\mathbb{E}^{Q}[X]\leq x_{0}\}
\end{equation*}%
for each $\gamma \in \Gamma $, using results \`{a} la \cite{FS16}. Third, we
take a minimum of $\gamma (X_{\gamma }^{\ast })$ over $\gamma \in \Gamma $,
to obtain the minimizer $\gamma ^{\ast }$. Finally, the position $X_{\gamma
^{\ast }}^{\ast }$ is a minimizer to the original problem \eqref{eq:opt3}.

The main point of this example is to show that optimization for star-shaped
risk measures is convenient if corresponding results on convex risk measures
are available.
\end{example}

\begin{remark}\label{rem:r2-1}
Recall that the convex risk measures $\gamma\in \Gamma$  in Theorem \ref{thm:conv}  
satisfy normalization, that is, $\gamma(0)=0$ for all $\gamma \in \Gamma$.  This simple property has a few implications.  Consider the problem of minimizing $\rho(\ell(a,\boldsymbol{S}))$ over $a\in A$ as in \eqref{eq:sopt1}, where $\rho=\min_{\gamma \in \Gamma} \gamma$. 
Suppose that there exists $a_0\in A$ representing a risk-free action  (e.g., perfect hedge, full insurance, non-defaultable bond purchase, or no participation, in different contexts), 
giving rise to $\ell (a_0,\boldsymbol{S})=c$ which is a constant. 
Our first observation is that, since each $\gamma $ is normalized, the problem can be rewritten as
 $$
 \min_{\gamma \in \Gamma}  \min_{a\in A} \Big( \gamma (\ell (a,\boldsymbol{S})) - \gamma (\ell(a_0,\boldsymbol{S}))\Big) +c.
 $$
That is, we can first optimize the relative improvement of our decision $a$ over the benchmark $a_0$ for each $\gamma$ and then look for the best improvement over all $\gamma\in \Gamma$. 
Second, if  $a \mapsto \ell (a,\boldsymbol{S})$  is bounded, then $ a \mapsto \gamma (\ell (a,\boldsymbol{S}))$ is uniformly bounded for $\gamma \in \Gamma$. 
Both the above observations may fail to hold if  $\gamma$ is not normalized.
In Section \ref{sec:conclusions}, we will briefly discuss   issues on normalization of $\gamma$. 
\end{remark}

\section{Law-invariant star-shaped risk measures}

\label{sec:law}


In this section, we discuss the important class of law-invariant risk
measures. Assume now that $\mathcal{X}=L^{\infty }\left( \Omega ,\mathcal{F}%
,P\right) $ where $P$ is an atomless probability measure. A risk measure $%
\rho $ is \emph{law-invariant} if two losses $X,Y\in \mathcal{X}$ that share
the same law under $P$ are equally risky:%
\begin{equation*}
X\overset{\mathrm{d}}{=}Y\implies \rho \left( X\right) =\rho \left( Y\right)
.
\end{equation*}

Most of risk measures that are used in the financial practice, like VaR and
ES, are law-invariant; we omit $P$ in $\mathrm{ES}^P_\alpha$ and $\mathrm{VaR}^P_\alpha$ in this
section. Law-invariance is the key property that facilitates statistical
inference for risk measures, thus converting distributional information into
risk assessment. For this reason, law-invariant risk measures are also
called \emph{statistical functionals}.

Since, according to Theorem \ref{thm:conv}, star-shaped risk measures can be
represented as minima of convex risk measures, one may naturally wonder
whether law-invariant and star-shaped risk measures are minima of
law-invariant and convex risk measures. Quite unexpectedly, the answer to
this question is no. Indeed, as shown by {Mao and Wang }\cite{MW20}, minima
of law-invariant and convex risk measures are consistent with respect to
second-order stochastic dominance ($\mathcal{SSD}$), and VaR ---while
law-invariant and star-shaped--- is not consistent with $\mathcal{SSD}$.

More specifically, we denote by $X\succcurlyeq _{\mathcal{SSD}}Y$ the fact
that $X\in \mathcal{X}$ second-order stochastically dominates $Y\in \mathcal{%
X}$, which means%
\begin{equation*}
\mathbb{E}_{P}[u(-X)]\geq \mathbb{E}_{P}[u(-Y)]
\end{equation*}%
for all increasing and concave $u:\mathbb{R}\rightarrow \mathbb{R}$ such
that the two expectations exist. 
Recall that our random variables describe losses, hence their negatives
describe gains.   A risk measure $\rho $ is said to be $\mathcal{SSD}$\emph{%
-consistent} if%
\begin{equation*}
X\succcurlyeq _{\mathcal{SSD}}Y\Longrightarrow \rho (X)\leq \rho (Y).
\end{equation*}%
In words, if less risky losses demand smaller capital reserves. The
following characterization is derived from the aforementioned \cite{MW20}.

\begin{theorem}
\label{thm:MaoWang} For a function $\rho :\mathcal{X}\rightarrow \mathbb{R}$%
, the following are equivalent:

\begin{enumerate}
\item[(i)] $\rho $ is a star-shaped and $\mathcal{SSD}$-consistent risk
measure;

\item[(ii)] there exists a star-shaped set $\mathcal{G}$ of increasing
functions $g:\left( 0,1\right) \rightarrow \mathbb{R}$ with $ g\left(
0+\right) \le  0$ such that%
\begin{equation}
\rho (X)=\inf_{g\in \mathcal{G}}\sup_{\alpha \in (0,1)}\{\mathrm{ES}_{\alpha
}(X)-g(\alpha )\}\qquad X\in \mathcal{X}.  \label{eq:es1}
\end{equation}
\end{enumerate}

Moreover, the class of star-shaped and $\mathcal{SSD}$-consistent risk measures is closed
under the operations considered in Theorem \ref{th:star_transf}.
\end{theorem}

The previous discussion on VaR, shows that the class of star-shaped and $%
\mathcal{SSD}$-consistent risk measures is strictly smaller than the class
of star-shaped and law-invariant ones. The characterization of this class of
risk measures concludes our analysis, by showing that star-shaped and
law-invariant risk measures are robustifications of VaR in the same sense in
which star-shaped and $\mathcal{SSD}$-consistent ones are robustifications
of ES.

\begin{theorem}
\label{prop:var} For a function $\rho :\mathcal{X}\rightarrow \mathbb{R}$,
the following are equivalent:

\begin{enumerate}
\item[(i)] $\rho $ is a star-shaped and law-invariant risk measure;

\item[(ii)] there exists a star-shaped set $\mathcal{G}$ of increasing
functions $g:\left( 0,1\right) \rightarrow \mathbb{R}$ with $g\left(
0+\right) \leq 0$ such that%
\begin{equation}
\rho (X)=\inf_{g\in \mathcal{G}}\sup_{\alpha \in (0,1)}\{\mathrm{VaR}%
_{\alpha }(X)-g(\alpha )\}\qquad X\in \mathcal{X}.  \label{eq:var1}
\end{equation}
\end{enumerate}

Moreover, the class of star-shaped and law-invariant risk measures is closed
under the operations considered in Theorem \ref{th:star_transf}.
\end{theorem}

According to Theorem \ref{thm:MaoWang}, there is a minimal element  in the set
of all star-shaped and $\mathcal{SSD}$-consistent risk measures dominating $\mathrm{VaR}_\alpha$.  
One may wonder whether this minimal element   coincides with $\mathrm{ES}_\alpha$, since $\mathrm{ES}_\alpha$ is known as the smallest law-invariant convex risk measure dominating $\mathrm{VaR}_\alpha$ (Proposition 5.4 of \cite{ADEH99} and Theorem 4.67 of \cite{FS16}).\footnote{We thank an anonymous referee for raising this question.}
The answer is affirmative, as stated in the following proposition, generalizing the aforementioned minimality of $\mathrm{ES}_\alpha$.
\begin{proposition}\label{prop:r1-1}
For $\alpha \in (0,1)$, the smallest $\mathcal{SSD}$-consistent risk measure dominating $\mathrm{VaR}_\alpha$ is $\mathrm{ES}_\alpha$.
\end{proposition}
In Proposition \ref{prop:r1-1},  star shapedness is not mentioned, but since
$\mathrm{ES}_\alpha$ is star-shaped, it is also the smallest star-shaped $\mathcal{SSD}$-consistent risk measure dominating $\mathrm{VaR}_\alpha$.

\section{Concluding remarks}

\label{sec:conclusions} This paper presents the first systematic
investigation of the minimalistic assumption of \emph{star shapedness} of
risk measures, highlighting the following features:

\begin{itemize}
\item star shapedness captures the equivalent rationality assumptions of 
\emph{increasing risk-to-exposure ratio} and \emph{increasing relative
ambiguity aversion};

\item star shapedness of risk measures is intimately linked to utility
functions that have \emph{local convexities} and possibly convex kinks;

\item star shapedness is the shared property of most risk measures adopted
in the financial practice and proposed in the literature, including \emph{%
Value-at-Risk} and its robustifications, on the one hand, and \emph{convex
risk measures} on the other;

\item star shapedness is preserved by \emph{competitive risk sharing} and 
\emph{robust aggregation} of risk measures, a property which is not enjoyed
by convex risk measures;

\item a risk measure is star-shaped if, and only if, it is a \emph{minimum
of (normalized) convex risk measures}, this makes its \emph{optimization tractable}.
\end{itemize}

The risk measures studied in this paper are static and real-valued
(unconditional). A promising direction of investigation is towards dynamic
and conditional risk measures. The natural extension of the definition of
star shapedness to $L^{0}$-valued risk measures presents itself as an
economically compelling and mathematically powerful one, sharing the gist
of subscale invariance of Bielecki, Cialenco, and Chen \cite{BCC15}.

 Aouani and Chateauneuf \cite{AC08}  characterized the class of distorted probabilities that  are exact capacities by means of probability distortions
$f: [0,1] \rightarrow \mathbb{R}$
that are star-shaped at both $0$ and $1$. Their Lemma 3.1 shows that these distortions are minima of convex distortions, thus, it has a ``flavor'' which is similar to our Theorem \ref{thm:conv}. But in this regard, it is worthwhile observing that the risk measures obtained by Choquet integration of these distorted probabilities are positively homogeneous (because they are comonotonic), and not only star-shaped. A gem, their Figure 1 shows that the iconic Friedman-Savage utility is star-shaped without being convex (see Section 
\ref{ex:fried} of this paper).

In this paper, all risk measures by definition satisfy normalization, a basic requirement for computing capital charges. If normalization is dropped from the definition of a risk measure, then every monotone and translation-invariant mapping is the minimum of (non-normalized) convex risk measures.\footnote{We thank an anonymous referee for raising this point.} This result was formalized in a recent study of Jia, Xia, and Zhao \cite{JXZ20}. 
The key difference between a star-shaped $\rho$ and a generic $\rho$ is normalization of the convex risk measures $\gamma\in\Gamma$ representing $\rho$ via $\rho=\min_{\gamma\in \Gamma} \gamma$.
We mention a few important differences made by normalization of $\gamma\in\Gamma$ (thus, star-shapedness of $\rho$). 
First, normalization is essential for the economic interpretation of the minimum representation, where each member $\gamma$ is interpreted a standalone risk measure (see Section \ref{sec:22}) or certainty equivalent (see Section \ref{sec:23}).
Second, normalization of $\gamma$ leads to a few non-trivial implications in optimization problems, as discussed in Remark \ref{rem:r2-1}.
Third, there are available tools for optimizing star-shaped functions in the literature; see e.g., 
  Rubinov  and Andramonov \cite{RA99}.

\subsection*{Acknowledgements} 
We thank an Editor and three anonymous referees for constructive comments, Simone Cerreia-Vioglio, Alain Chateauneuf, Paul Embrechts, Tiantian Mao,   Massimo Marinacci,   Liang Peng, and Qihe Tang for very helpful  discussions,
Giulio Principi and Paolo Leonetti for brilliant research assistance, 
as well as the PRIN grant (2017CY2NCA, 2017TA7TYC) and the NSERC grants (RGPIN-2018-03823,
RGPAS-2018-522590) for financial support.\\
\textbf{Correspondence.} 
Correspondence should be addressed to R.~Wang (\texttt{wang@uwaterloo.ca}).

\begin{APPENDIX}{Proofs}

\noindent \textbf{Proof of Proposition \ref{pro:star_char}.} \underline{(i) implies
(ii)}. If $\alpha \in \left( 0,1\right) $, then $1/\alpha \in \left( 1,\infty
\right) $, and, for all $X\in \mathcal{X}$, star shapedness implies%
\begin{equation*}
\rho \left( X\right) =\rho \left( \frac{1}{\alpha }\left( \alpha X\right)
\right) \geq \frac{1}{\alpha }\rho \left( \alpha X\right) .
\end{equation*}%
\underline{(ii) implies (iii)}. Let $\beta >\gamma >0$, then, for all $X\in \mathcal{X}$%
, $\gamma /\beta \in \left( 0,1\right) $ implies%
\begin{equation*}
\rho \left( \gamma X\right) =\rho \left( \frac{\gamma }{\beta }\left( \beta
X\right) \right) \leq \frac{\gamma }{\beta }\rho \left( \beta X\right)
\end{equation*}%
that is, $\rho \left( \beta X\right) /\beta \geq \rho \left( \gamma X\right)
/\gamma $.

\underline{(iii) implies (i)}. For all $X\in \mathcal{X}$ and all $\lambda >1$, since $%
r_{X}$ is increasing, one has 
\begin{equation*}
\frac{\rho (\lambda X)}{\lambda }\geq \frac{\rho (1X)}{1}=\rho (X)
\end{equation*}%
as wanted.\hfill $\blacksquare \bigskip $

\noindent \textbf{Proof of Proposition \ref{pro:star_acc}.} \underline{(i) implies
(ii)}. Let $X\in \mathcal{A}_{\rho }$ and $\alpha \in \left( 0,1\right) $. By
Proposition \ref{pro:star_char}, $\rho \left( \alpha X\right) \leq \alpha
\rho \left( X\right) \leq \alpha 0=0$, that is, $\alpha X\in \mathcal{A}%
_{\rho }$. Normalization implies that the same happens for $\alpha =0$,
while the fact that $1X=X$ covers the case $\alpha =1$.

\underline{(ii) implies (iii)}. In fact, $\mathcal{A}_{\rho }$ is an acceptance set, it is
star-shaped, and by \eqref{eq:tri} we have $\rho =\rho _{\mathcal{A}
_{\rho }}$.

\underline{(iii) implies (i)}. Let $X\in \mathcal{X}$ and $\alpha \in \left( 0,1\right) $%
. If $m\in \mathbb{R}$ and $X-m\in \mathcal{A}$, then star shapedness of $%
\mathcal{A}$ implies $\alpha X-\alpha m\in \mathcal{A}$, thus 
\begin{equation*}
\rho \left( \alpha X\right) =\inf \left\{ k\in \mathbb{R}\mid \alpha X-k\in 
\mathcal{A}\right\} \leq \alpha m.
\end{equation*}%
Since $m$ was arbitrarily chosen, one has 
\begin{equation*}
\frac{\rho \left( \alpha X\right) }{\alpha }\leq \inf \left\{ m\in \mathbb{R}
\mid X-m\in \mathcal{A}\right\} =\rho \left( X\right) .
\end{equation*}%
By Proposition \ref{pro:star_char}, $\rho $ is star-shaped. \hfill $%
\blacksquare \bigskip $

\noindent \textbf{Proof of Proposition \ref{pro:homo_super}.} 
\underline{(i) implies
(ii)}. Let $X\in \mathcal{X}$. We know from Proposition \ref{pro:star_char}
that $r_{X}:\beta \mapsto \rho (\beta X)/\beta $ is increasing on $(0,\infty
)$. Subadditivity of $\rho $ implies that%
\begin{equation*}
\rho (2^{n}X)=\rho (2^{n-1}X+2^{n-1}X)\leq 2\rho (2^{n-1}X)
\end{equation*}%
for all $n\in \mathbb{Z}$, and hence 
\begin{equation*}
r_{X}(2^{n})=\frac{\rho (2^{n}X)}{2^{n}}\leq \frac{\rho (2^{n-1}X)}{2^{n-1}}
=r_{X}(2^{n-1})\leq r_{X}(2^{n}).
\end{equation*}%
Therefore, the increasing function $r_{X}$ is constant on the set $%
\{2^{n}:n\in \mathbb{Z}\}$, and this forces it to be constant on $(0,\infty
) $, showing positive homogeneity.

The rest is straightforward. \hfill $\blacksquare \bigskip$

\noindent \textbf{Proof of Theorem \ref{th:star_transf}. }It is sufficient
to prove the theorem for Choquet averages and inf-convolutions, since the
supremum and infimum operations are special cases of the former.\medskip

\noindent \textbf{Choquet averages.} Let $\mu $ be a capacity on the set of
all parts of $I$. Note that, for each $X\in \mathcal{X}$, 
\begin{equation*}
\begin{array}{llll}
\varrho ^{X}: & I & \rightarrow & \mathbb{R} \\ 
& i & \mapsto & \varrho ^{X}\left( i\right) =\rho _{i}\left( X\right)%
\end{array}%
\end{equation*}%
is such that $\inf_{\omega \in \Omega }X\left( \omega \right) \leq \rho
_{i}\left( X\right) \leq \sup_{\omega \in \Omega }X\left( \omega \right) $
for all $i\in I$, thus $\varrho ^{X}$ is a bounded function from $I$ to $%
\mathbb{R}$. Denote by $B\left( I\right) $ the set of all bounded functions
from $I$ to $\mathbb{R}$ and observe that%
\begin{equation*}
\begin{array}{llll}
\mu : & B\left( I\right) & \rightarrow & \mathbb{R} \\ 
& f & \mapsto & \mu \left( f\right) =\int_{I}f\left( i\right) \,\mathrm{d}
\mu \left( i\right)%
\end{array}%
\end{equation*}%
is a positively homogeneous risk measure on $B ( I ) $.

Next we show that if $\left\{ \rho _{i}\right\} _{i\in I}$ is a collection
of star-shaped risk measures, then%
\begin{equation*}
\rho _{\mu }(X)=\int_{I}\rho _{i}(X)\,\mathrm{d}\mu \left( i\right) =\mu
\left( \varrho ^{X}\right) \qquad X\in \mathcal{X}
\end{equation*}%
is a star-shaped risk measure too.\medskip

\noindent 1. \emph{Monotonicity:} if $X\geqq Y$, then $\rho _{\mu }(X)\geq
\rho _{\mu }(Y)$.

In fact, $X\geqq Y$ implies $\rho _{i}(X)\geq \rho _{i}(Y)$\ for all $i\in I$%
, that is, $\varrho ^{X}\geqq \varrho ^{Y}$, whence 
\begin{equation*}
\rho _{\mu }(X)=\mu \left( \varrho ^{X}\right) \geq \mu \left( \varrho
^{Y}\right) =\rho _{\mu }(Y).
\end{equation*}

\noindent 2. \emph{Translation invariance:} $\rho _{\mu }(X-m)=\rho _{\mu
}(X)-m$ for all $X\in \mathcal{X}$ and all $m\in \mathbb{R}$.

Given any $X\in \mathcal{X}$ and any $m\in \mathbb{R}$,%
\begin{equation*}
\varrho ^{X-m}\left( i\right) =\rho _{i}(X-m)=\rho _{i}(X)-m=\varrho
^{X}\left( i\right) -m\qquad i\in I
\end{equation*}%
that is, $\varrho ^{X-m}=\varrho ^{X}-m$, whence 
\begin{equation*}
\rho _{\mu }(X-m)=\mu \left( \varrho ^{X-m}\right) =\mu \left( \varrho
^{X}-m\right) =\mu \left( \varrho ^{X}\right) -m=\rho _{\mu }(X)-m.
\end{equation*}

\noindent 3. \emph{Normalization:} $\rho _{\mu }(0)=0$.

Clearly $\varrho ^{0}\left( i\right) =\rho _{i}\left( 0\right) =0$ for all $%
i\in I$, that is, $\varrho ^{0}\equiv 0$, whence%
\begin{equation*}
\rho _{\mu }\left( 0\right) =\mu \left( \varrho ^{0}\right) =\mu \left(
0\right) =0.
\end{equation*}%
\noindent 4. \emph{Star shapedness:} $\rho _{\mu }(\lambda X)\geq \lambda
\rho _{\mu }(X)$ for all $X\in \mathcal{X}$ and all $\lambda >1$.\medskip

Given any $X\in \mathcal{X}$ and any $\lambda >1$, $\rho _{i}(\lambda X)\geq
\lambda \rho _{i}(X)$\ for all $i\in I$, that is, $\varrho ^{\lambda X}\geqq
\lambda \varrho ^{X}$, whence 
\begin{equation*}
\rho _{\mu }(\lambda X)=\mu \left( \varrho ^{\lambda X}\right) \geq \mu
\left( \lambda \varrho ^{X}\right) =\lambda \mu \left( \varrho ^{X}\right)
=\lambda \rho _{\mu }(X).
\end{equation*}%
\medskip

\noindent \textbf{Infimal convolutions.} Set $I=\left\{ 1,2,...,n\right\} $.
For each $X\in \mathcal{X}$ define the sets%
\begin{eqnarray*}
\mathcal{D}\left( X\right) &=&\left\{ \boldsymbol{Y}\in \mathcal{X}
^{I}\mid \sum_{i\in I}Y_{i}=X\right\} \\
\mathcal{S}\left( X\right) &=&\left\{ \boldsymbol{Y}\in \mathcal{X}
^{I}\mid \sum_{i\in I}Y_{i}\geqq X\right\}
\end{eqnarray*}%
and note that $\mathcal{D}\left( X\right) \subseteq \mathcal{S}\left(
X\right) \subseteq \mathcal{S}\left( Y\right) $ if $Y$ in $\mathcal{X}$ is
such that $Y\leqq X$. Consider the function%
\begin{equation*}
\begin{array}{llll}
\boldsymbol{\rho }: & \mathcal{X}^{I} & \rightarrow & \mathbb{R} \\ 
& \boldsymbol{Y} & \mapsto & \boldsymbol{\rho }\left( \boldsymbol{Y}\right)
=\sum\limits_{i\in I}\rho _{i}\left( Y_{i}\right).%
\end{array}%
\end{equation*}%
Note that%
\begin{equation*}
\rho _{\diamond }(X)=\inf \left\{ \sum_{i\in I}\rho _{i}\left( Y_{i}\right)
\ \left\vert \ Y_{i}\in \mathcal{X}\text{ for all }i\in I\text{ and\ }
\sum_{i\in I}Y_{i}=X\right. \right\} =\inf_{\boldsymbol{Y}\in \mathcal{D}
\left( X\right) }\boldsymbol{\rho }\left( \boldsymbol{Y}\right)
\end{equation*}%
for all $X\in \mathcal{X}$. Next we show that 
\begin{equation*}
\rho _{\diamond }(X)=\inf_{\boldsymbol{Y}\in \mathcal{S}\left( X\right) } 
\boldsymbol{\rho }\left( \boldsymbol{Y}\right) .
\end{equation*}%
Since $\mathcal{D}\left( X\right) \subseteq \mathcal{S}\left( X\right) $,
one has $\rho _{\diamond }(X)=\inf_{\boldsymbol{Y}\in \mathcal{D}\left(
X\right) }\boldsymbol{\rho }\left( \boldsymbol{Y}\right) \geq \inf_{ 
\boldsymbol{Y}\in \mathcal{S}\left( X\right) }\boldsymbol{\rho }\left( 
\boldsymbol{Y}\right) $. As to the converse inequality, observe that, for
each $\boldsymbol{Y}\in \mathcal{S}\left( X\right) $, we have that $%
Z=X-\sum_{i\in I}Y_{i}\leqq 0$ and $Z\in \mathcal{X}$, and we can define $%
\bar{Y}_{1}=Y_{1}+Z\leqq Y_{1}$. With this $\left( \bar{Y}%
_{1},Y_{2},\dots ,Y_{n}\right) \in \mathcal{X}^{I}$ and $\bar{Y}%
_{1}+Y_{2}+\dots+Y_{n}=Y_{1}+Z+Y_{2}+\dots+Y_{n}=X$, thus $\left( \bar{Y}%
_{1},Y_{2},\dots ,Y_{n}\right) \in \mathcal{D}\left( X\right) $ and%
\begin{equation*}
\rho _{\diamond }\left( X\right) =\inf_{\boldsymbol{V}\in \mathcal{D}\left(
X\right) }\boldsymbol{\rho }\left( \boldsymbol{V}\right) \leq \boldsymbol{\
\rho }\left( \bar{Y}_{1},Y_{2},\dots,Y_{n}\right) =\rho _{1}\left( \bar{Y}
_{1}\right) +\dots+\rho _{n}\left( Y_{n}\right) \leq \sum_{i\in I}\rho
_{i}\left( Y_{i}\right) =\boldsymbol{\rho }\left( \boldsymbol{Y}\right) .
\end{equation*}%
Then $\rho _{\diamond }\left( X\right) $ is a lower bound for $\left\{ 
\boldsymbol{\rho }\left( \boldsymbol{Y}\right) \mid  \boldsymbol{Y}\in \mathcal{S}%
\left( X\right) \right\} $ hence $\rho _{\diamond }(X)\leq \inf_{\boldsymbol{%
\ Y}\in \mathcal{S}\left( X\right) }\boldsymbol{\rho }\left( \boldsymbol{Y}%
\right) $.\medskip

\noindent 1. \emph{Monotonicity:} if $X\geqq Y$, then $\rho _{\diamond
}(X)\geq \rho _{\diamond }(Y)$.

As observed, $X\geqq Y$ implies $\mathcal{S}\left( X\right) \subseteq 
\mathcal{S}\left( Y\right) $, whence%
\begin{equation*}
\rho _{\diamond }(X)=\inf_{\boldsymbol{V}\in \mathcal{S}\left( X\right) } 
\boldsymbol{\rho }\left( \boldsymbol{V}\right) \geq \inf_{\boldsymbol{V}\in 
\mathcal{S}\left( Y\right) }\boldsymbol{\rho }\left( \boldsymbol{V}\right)
=\rho _{\diamond }(Y).
\end{equation*}

\noindent 2. \emph{Translation invariance:} $\rho _{\diamond }(X-m)=\rho
_{\diamond }(X)-m$ for all $X\in \mathcal{X}$ and all $m\in \mathbb{R}$.

Given any $X\in \mathcal{X}$ and any $m\in \mathbb{R}$, if $\boldsymbol{Y}%
\in \mathcal{D}\left( X-m\right) $, then $\left(
Y_{1}+m,Y_{2},...,Y_{n}\right) \in \mathcal{D}\left( X\right) $, thus%
\begin{equation*}
\rho _{\diamond }\left( X\right) \leq \boldsymbol{\rho }\left(
Y_{1}+m,Y_{2},...,Y_{n}\right) =\rho _{1}\left( Y_{1}+m\right) +...+\rho
_{n}\left( Y_{n}\right)
\end{equation*}%
and 
\begin{equation*}
\rho _{\diamond }\left( X\right) -m\leq \rho _{1}\left( Y_{1}\right)
+...+\rho _{n}\left( Y_{n}\right) =\boldsymbol{\rho }\left( \boldsymbol{Y}
\right) .
\end{equation*}%
Hence 
\begin{equation*}
\rho _{\diamond }\left( X\right) -m\leq \inf_{\boldsymbol{Y}\in \mathcal{D}
\left( X-m\right) }\boldsymbol{\rho }\left( \boldsymbol{Y}\right) =\rho
_{\diamond }\left( X-m\right)
\end{equation*}%
that is, $\rho _{\diamond }\left( X\right) -m\leq \rho _{\diamond }\left(
X-m\right) $ for all $\left( X,m\right) \in \mathcal{X}\times \mathbb{R}$.
But then, replacing $m$ with $-m$, and $X$ with $X-m$,%
\begin{equation*}
\rho _{\diamond }\left( X-m\right) +m=\rho _{\diamond }\left( X-m\right)
-(-m)\leq \rho _{\diamond }\left( \left[ X-m\right] -(-m)\right) =\rho
_{\diamond }\left( X\right)
\end{equation*}%
that is, $\rho _{\diamond }\left( X\right) -m\geq \rho _{\diamond }\left(
X-m\right) $ for all $\left( X,m\right) \in \mathcal{X}\times \mathbb{R}$.

\noindent 3. \emph{Normalization:} $\rho _{\diamond }(0)=0$.

Clearly, 
\begin{equation*}
\rho _{\diamond }\left( 0\right) =\inf \left\{ \sum_{i\in I}\rho _{i}\left(
Y_{i}\right) \ \left\vert \ Y_{i}\in \mathcal{X}\text{ for all }i\in I\text{
and\ }\sum_{i\in I}Y_{i}=0\right. \right\} \leq \sum_{i\in I}\rho _{i}\left(
0\right) =0.
\end{equation*}%
Then $\rho _{\diamond }\left( 0\right) =0$, if, and only if,%
\begin{equation*}
\inf \left\{ \sum_{i\in I}\rho _{i}\left( Y_{i}\right) \ \left\vert \
Y_{i}\in \mathcal{X}\text{ for all }i\in I\text{ and\ }\sum_{i\in
I}Y_{i}=0\right. \right\} \geq 0
\end{equation*}%
that is, if, and only if,%
\begin{equation*}
\sum_{i\in I}\rho _{i}\left( Y_{i}\right) \geq 0
\end{equation*}%
for all $Y_{1},Y_{2},...,Y_{n}\in \mathcal{X}$ such that $\sum_{i\in
I}Y_{i}=0$, which is precisely condition \eqref{eq:split}.\medskip

\textbf{Remark.} Observe that $\rho _{\diamond }(X)=\inf_{\boldsymbol{Y}
\in \mathcal{D}\left( X\right) }\boldsymbol{\rho }\left( \boldsymbol{Y}
\right) $ guarantees that $\rho _{\diamond }(X)\in \left[ -\infty ,\infty
\right) $ for all $X\in \mathcal{X}$. But together with translation
invariance normalization implies that $\rho _{\diamond }(m)=m$ for all $m\in 
\mathbb{R}$, then monotonicity of $\rho _{\diamond }$ and boundedness of the
elements of $\mathcal{X}$ guarantee that $\rho _{\diamond }(X)\in \mathbb{R}$
for all $X\in \mathcal{X}$.
\medskip

\noindent 4. \emph{Star shapedness:} $\rho _{\diamond }(\lambda X)\geq
\lambda \rho _{\diamond }(X)$ for all $X\in \mathcal{X}$ and all $\lambda >1$%
.\medskip

Given any $X\in \mathcal{X}$ and any $\lambda >0$, note that 
\begin{equation*}
\mathcal{D}\left( \lambda X\right) =\lambda \mathcal{D}\left( X\right) .
\end{equation*}%
In fact, $\boldsymbol{Y}\in \mathcal{D}\left( \lambda X\right) $ implies $%
\boldsymbol{Y}\in \mathcal{X}^{I}$ and $\sum_{i\in I}Y_{i}=\lambda X$, then $%
\boldsymbol{Z}=\lambda ^{-1}\boldsymbol{Y}\in \mathcal{X}^{I}$ and $%
\sum_{i\in I}Z_{i}=\sum_{i\in I}\lambda ^{-1}Y_{i}=\lambda ^{-1}\sum_{i\in
I}Y_{i}=X$, thus $\boldsymbol{Z}\in \mathcal{D}\left( X\right) $ and $%
\boldsymbol{Y}=\lambda \boldsymbol{Z}$. This shows that $\mathcal{D}\left(
\lambda X\right) \subseteq \lambda \mathcal{D}\left( X\right) $ for all $%
\left( X,\lambda \right) \in \mathcal{X}\times \left( 0,\infty \right) $,
but then 
\begin{equation*}
\lambda \mathcal{D}\left( X\right) =\lambda \mathcal{D}\left( \frac{1}{
\lambda }\left( \lambda X\right) \right) \subseteq \lambda \left( \frac{1}{
\lambda }\mathcal{D}\left( \lambda X\right) \right) =\mathcal{D}\left(
\lambda X\right)
\end{equation*}%
and equality holds. Therefore, if $\lambda >1$,%
\begin{align*}
\rho _{\diamond }(\lambda X)& =\inf_{\boldsymbol{Y}\in \mathcal{D}\left(
\lambda X\right) }\sum_{i\in I}\rho _{i}\left( Y_{i}\right) =\inf_{ 
\boldsymbol{V}\in \mathcal{D}\left( X\right) }\sum_{i\in I}\rho _{i}\left(
\lambda V_{i}\right) \\
& \geq \inf_{\boldsymbol{V}\in \mathcal{D}\left( X\right) }\sum_{i\in
I}\lambda \rho _{i}\left( V_{i}\right) =\lambda \inf_{\boldsymbol{V}\in 
\mathcal{D}\left( X\right) }\sum_{i\in I}\rho _{i}\left( V_{i}\right)
=\lambda \rho _{\diamond }(X)
\end{align*}%
as wanted.\hfill $\blacksquare \bigskip $

\noindent \textbf{Proof of Theorem \ref{thm:conv}.} Since positively
homogeneous risk measures are obviously star-shaped, the statement says that
they admit both representations:%
\begin{equation*}
\rho (X)=\min \left\{ \gamma (X)\mid  \gamma \geqq \rho \text{ is a convex risk
measure}\right\}
\end{equation*}%
and 
\begin{equation*}
\rho (X)=\min \left\{ \delta (X)\mid  \delta \geqq \rho \text{ is a coherent risk
measure}\right\} \text{.}
\end{equation*}%
In this proof, when we write \textquotedblleft star-shaped (resp.~positively
homogeneous),\textquotedblright\ we mean \textquotedblleft to obtain the
general result that holds for all star-shaped risk measures (resp.~to obtain
the special result that holds for the positively homogeneous
ones).\textquotedblright

\underline{(i) implies (ii)}. For each $Y\in \mathcal{X}$ set 
\begin{equation*}
\mathcal{A}_{Y}=\mathrm{co}\{Y-\rho (Y),0\} -\mathcal{X}^{+}
\end{equation*}%
where $\mathcal{X}^{+}$ is the cone of non-negative  elements in $\mathcal{X}$,
and $\mathrm{co}\{Y-\rho (Y),0\}$ is the convex set (resp.~the convex cone) 
generated by $\{Y-\rho (Y),0\}$. 
Recall that the \emph{convex cone} of a subset $S$ of a vector space is the
set of all elements of the form $a s+b t $
such that $a, b \in \mathbb{R}_{+}$ and $s,t\in S$. 
We note that $\mathcal A_Y$ is a convex set (resp.~a convex
cone), because it is a sum of convex sets (resp.~convex cones).

Since $\rho $ is translation invariant,  $\rho(Y-\rho(Y))=0$ and $Y-\rho (Y)\in \mathcal{A}_{\rho }$. 

\begin{itemize}
\item In the star-shaped case, by Proposition \ref{pro:star_acc}, $\mathcal{%
A }_{\rho }$ is star-shaped. For any $\alpha \in \left[ 0,1\right] $ we
have $\alpha (Y-\rho (Y)) \in \mathcal A_\rho$, implying $\alpha (Y-\rho (Y))-\mathcal{X}^{+}\subseteq \mathcal{A}_{\rho }$.
Hence,
$
\mathcal A_Y\subseteq \mathcal A_\rho.
$  
\item In the positively homogeneous case, $\mathcal{%
A }_{\rho }$ is conic. For any $\alpha \in  [ 0,\infty) $ we
have $\alpha (Y-\rho (Y)) \in \mathcal A_\rho$, implying $\alpha (Y-\rho (Y))-\mathcal{X}^{+}\subseteq \mathcal{A}_{\rho }$.
Hence,
$
\mathcal A_Y\subseteq \mathcal A_\rho.
$  
\end{itemize} 


Since $0\in \mathrm{co}%
\{Y-\rho (Y),0\}$ then $0\in \mathcal{A}_{Y}$, it
follows that%
\begin{equation*}
	0\leq \sup \left\{ m\in \mathbb{R}\mid m\in \mathcal{A}_{Y}\right\} \leq
	\max \left\{ m\in \mathbb{R}\mid m\in \mathcal{A}_{\rho }\right\} =0,
\end{equation*}%
and so $0=\max \left\{ m\in \mathbb{R}\mid m\in \mathcal{A}_{Y}\right\} $.
Moreover, it is straightforward to check that $ Z\leqq X$ 
implies  $Z\in \mathcal A_Y$ if   $X\in \mathcal{A}_{Y}$.

So far, we know that $\mathcal{A}_{Y}$ is a convex (resp.~coherent)
acceptance set included in $\mathcal{A}_{\rho }$.
\medskip

Now, notice that since $Y-\rho (Y)\in \mathcal{A}_{Y}$, we have that 
\begin{equation*}
\rho _{\mathcal{A}_{Y}}(Y)=\inf \{m\in \mathbb{R}\mid Y-m\in \mathcal{A}
_{Y}\}\leq \rho (Y).
\end{equation*}%
But since $\mathcal{A}_{Y}\subseteq \mathcal{A}_{\rho }$, then, for all $%
X\in \mathcal{X}$,%
\begin{eqnarray*}
\rho (X) &=&\inf \{m\in \mathbb{R}\mid X-m\in \mathcal{A}_{\rho }\} \\
&\leq &\inf \{m\in \mathbb{R}\mid X-m\in \mathcal{A}_{Y}\}=\rho _{\mathcal{A}
_{Y}}(X),
\end{eqnarray*}%
and in particular, $\rho (Y)=\rho _{_{\mathcal{A}_{Y}}}(Y)$.

Summing up, for all $X\in \mathcal{X}$,
\begin{align*}
\rho (X)  \leq  \inf_{Y\in \mathcal{X}}\rho _{\mathcal{A}_{Y}}(X) \quad \mbox{~and~}\quad
\rho (X)  = \rho _{\mathcal{A}_{X}}(X) 
\end{align*}%
that is%
\begin{equation*}
\rho (X)=\min_{Y\in \mathcal{X}}\rho _{\mathcal{A}_{Y}}(X).
\end{equation*}%
\emph{A fortiori}, letting $\Gamma $ be the set of all convex (resp.~coherent) risk measures dominating $\rho $, since $\rho _{\mathcal{A}_{X}}$
belongs to $\Gamma $, 
\begin{align*}
\rho (X)  \leq  \inf_{\gamma \in \Gamma }\gamma (X) \quad \mbox{~and~}\quad
\rho (X)  = \rho _{\mathcal{A}_{X}}(X)  
\end{align*}%
that is 
\begin{equation*}
\rho (X)=\min_{\gamma \in \Gamma }\gamma (X).
\end{equation*}%
This completes the main implication of the theorem and shows that $\Gamma $
can be chosen as the set of all convex (resp.~coherent) risk measures
dominating $\rho $.

\underline{(ii) implies (iii)}. Let $\Gamma $ be a collection of convex (resp.~coherent)
risk measures such that%
\begin{equation*}
\rho (X)=\min_{\gamma \in \Gamma }\gamma (X)\qquad X\in \mathcal{X}
\end{equation*}%
and denote by $\mathcal{A}_{\gamma }$ the acceptance set of each $\gamma \in
\Gamma $. Theorem \ref{th:star_transf} guarantees that $\rho $ is a
star-shaped risk measure, moreover,%
\begin{eqnarray*}
\mathcal{A}_{\rho } &=&\left\{ X\in \mathcal{X}\mid \rho \left( X\right)
\leq 0\right\} =\left\{ X\in \mathcal{X}\mid \gamma \left( X\right) \leq 0 
\text{ for some }\gamma \in \Gamma \right\} \\
&=&\left\{ X\in \mathcal{X}\mid X\in \mathcal{A}_{\gamma }\text{ for some }
\gamma \in \Gamma \right\} =\bigcup_{\gamma \in \Gamma }\mathcal{A}_{\gamma
}.
\end{eqnarray*}%
Then $\left\{ \mathcal{A}_{\gamma }\right\} _{\gamma \in \Gamma }$ is a
family of convex (resp.~coherent) acceptance sets such that%
\begin{eqnarray*}
\rho (X) &=&\min \left\{ m\in \mathbb{R}\mid X-m\in \mathcal{A}_{\rho
}\right\} =\min \left\{ m\in \mathbb{R}\ \left\vert \ X-m\in \bigcup_{\gamma
\in \Gamma }\mathcal{A}_{\gamma }\right. \right\} \\
&=&\min \left\{ m\in \mathbb{R}\mid X-m\in \mathcal{A}_{\gamma }\text{ for
some }\gamma \in \Gamma \right\}
\end{eqnarray*}%
for all $X\in \mathcal{X}$. Clearly, when $\Gamma $ is the set of all convex
(resp.~coherent) risk measures dominating $\rho $, $\left\{ \mathcal{A}%
_{\gamma }\right\} _{\gamma \in \Gamma }$ is the family of their acceptance
sets.

\underline{(iii) implies (i)}. Let $\left\{ \mathcal{A}_{\beta }\right\} _{\beta \in B}$
be a family of convex (resp.~coherent) acceptance sets such that%
\begin{eqnarray*}
\rho (X) &=&\min \left\{ m\in \mathbb{R}\mid X-m\in \mathcal{A}_{\beta } 
\text{ for some }\beta \in B\right\} \\
&=&\min \left\{ m\in \mathbb{R}\ \left\vert \ X-m\in \bigcup_{\beta \in B} 
\mathcal{A}_{\beta }\right. \right\} .
\end{eqnarray*}%
For each $X\in \mathcal{X}$ define 
\begin{equation*}
\begin{array}{llll}
f_{X}: & \mathbb{R} & \rightarrow & \mathcal{X} \\ 
& m & \mapsto & X-m%
\end{array}%
\end{equation*}%
and note that%
\begin{equation*}
\left. X-m\in \bigcup_{\beta \in B}\mathcal{A}_{\beta }\right. \iff
f_{X}\left( m\right) \in \bigcup_{\beta \in B}\mathcal{A}_{\beta }\iff m\in
f_{X}^{-1}\left( \bigcup_{\beta \in B}\mathcal{A}_{\beta }\right) \iff m\in
\bigcup_{\beta \in B}f_{X}^{-1}\left( \mathcal{A}_{\beta }\right) .
\end{equation*}%
With this%
\begin{eqnarray*}
\rho (X) &=&\inf \left\{ m\in \mathbb{R}\ \left\vert \ m\in \bigcup_{\beta
\in B}f_{X}^{-1}\left( \mathcal{A}_{\beta }\right) \right. \right\} =\inf
\left( \bigcup_{\beta \in B}f_{X}^{-1}\left( \mathcal{A}_{\beta }\right)
\right) \\
&=&\inf_{\beta \in B}\left\{ \inf f_{X}^{-1}\left( \mathcal{A}_{\beta
}\right) \right\} =\inf_{\beta \in B}\left\{ \inf \left\{ m\in \mathbb{R}
\mid m\in f_{X}^{-1}\left( \mathcal{A}_{\beta }\right) \right\} \right\} \\
&=&\inf_{\beta \in B}\underset{\rho _{\beta }\left( X\right) }{\underbrace{
\inf \left\{ m\in \mathbb{R}\mid X-m\in \mathcal{A}_{\beta }\right\} }}
\end{eqnarray*}%
where $\rho _{\beta }$ is the risk measure generated by $\mathcal{A}_{\beta
} $, which is convex (resp.~coherent) since the $\mathcal{A}_{\beta }$'s are
convex (resp.~coherent) acceptance sets. But then Theorem \ref%
{th:star_transf} guarantees that $\rho $ is a star-shaped risk measure (if,
in addition, the $\rho _{\beta }$'s are coherent, immediate verification
yields positive homogeneity of $\rho $).\hfill $\blacksquare \bigskip $

\noindent \textbf{Proof of Theorem \ref{th:rep-opre}.} By Theorem \ref%
{th:star_transf}, we know that the Choquet average, supremum, infimum, and
inf-convolution are star-shaped risk measures. We proceed to show that
representations  \eqref{eq:op1}-\eqref{eq:op4} hold for an arbitrary $X\in 
\mathcal{X}$. For ease of notation, set $\boldsymbol{\Gamma }=\prod_{i\in
I}\Gamma _{i}$ and denote by $\boldsymbol{\gamma }=\left( \gamma _{i}\right)
_{i\in I}$ its generic element, and note that%
\begin{equation*}
\int \Gamma _{i}\,\mathrm{d}\mu \left( i\right) =\left\{ \int_{I}\gamma
_{i}\,\mathrm{d}\mu \left( i\right) \ \left\vert \ \boldsymbol{\gamma }\in 
\boldsymbol{\Gamma }\right. \right\} \text{\quad and\quad }\square _{i\in
I}\Gamma _{i}=\left\{ \square _{i\in I}\gamma _{i}\mid \boldsymbol{\gamma }
\in \boldsymbol{\Gamma }\right\} .
\end{equation*}%
$\medskip $

\noindent \textbf{Choquet averages.} For each $i\in I$, choose $\gamma
_{i}^{\ast }\in \Gamma _{i}$ such that $\gamma _{i}^{\ast }(X)=\rho _{i}(X)$%
, so that 
\begin{equation*}
\inf_{\boldsymbol{\gamma }\in \boldsymbol{\Gamma }}\int_{I}\gamma _{i}(X)\, 
\mathrm{d}\mu \left( i\right) \leq \int_{I}\gamma _{i}^{\ast }(X)\,\mathrm{d}
\mu \left( i\right) =\int_{I}\rho _{i}(X)\,\mathrm{d}\mu \left( i\right)
=\int_{I}\min_{\gamma _{i}\in \Gamma _{i}}\gamma _{i}(X)\,\mathrm{d}\mu
\left( i\right) \leq \inf_{\boldsymbol{\gamma }\in \boldsymbol{\Gamma }
}\int_{I}\gamma _{i}(X)\,\mathrm{d}\mu \left( i\right) .
\end{equation*}%
%
%
%
%
%
%
%
%
%
%
%
%
%
%
%
%
%
%
%
%
%
%
%
%
%
%
%
%
%
%
%
%
%
%
%
%
%
%
%
%
%
%
%
%
%
%
%
%
%
%
%
%
%
%
%
%
%
%
%
%
%
%
%
%
%
%
%
%
%
%
%
%
%
%
%
%
%
%
%
%
%
%
%
%
%
%
%
%
%
Since $\boldsymbol{\gamma }^{\ast }\in \boldsymbol{\Gamma }$, one has 
\begin{equation*}
\int_{I}\rho _{i}(X)\,\mathrm{d}\mu \left( i\right) =\int_{I}\gamma _{i}^{\ast }(X)\,\mathrm{%
\ d}\mu \left( i\right) =\min_{\boldsymbol{\gamma }\in \boldsymbol{\Gamma }
}\int_{I}\gamma _{i}(X)\,\mathrm{d}\mu \left( i\right) =\min_{\delta \in
\int \Gamma _{i}\,\mathrm{d}\mu \left( i\right) }\delta \left( X\right) .
\end{equation*}%
\medskip

\noindent \textbf{Suprema.} Since 
\begin{equation*}
\sup_{i\in I}\rho _{i}(X)
\end{equation*}%
is a star-shaped risk measure, by Theorem \ref{thm:conv},%
\begin{eqnarray*}
\sup_{i\in I}\rho _{i}(X) &=&\min \left\{ \gamma \left( X\right) \mid \gamma 
\text{ is a convex risk measure and, }\forall Y\in \mathcal{X},\text{ }
\gamma \left( Y\right) \geq \sup_{i\in I}\rho _{i}\left( Y\right) \right\} \\
&=&\min \left\{ \gamma \left( X\right) \mid \gamma \text{ is a convex risk
measure and }\forall Y\in \mathcal{X},\text{ }\forall i\in I,\ \gamma \left(
Y\right) \geq \rho _{i}\left( Y\right) \right\} \\
&=&\min \left\{ \gamma \left( X\right) \mid \gamma \text{ is a convex risk
measure and }\forall i\in I,\ \forall Y\in \mathcal{X},\ \gamma \left(
Y\right) \geq \rho _{i}\left( Y\right) \right\} \\
&=&\min \left\{ \gamma \left( X\right) \mid \gamma \text{ is a convex risk
measure and }\forall i\in I,\ \gamma \in \tilde{\Gamma}_{i}\right\}.
\end{eqnarray*}%
Moreover, since each $\tilde{\Gamma}_{i}$ consists of convex risk measures,
one has%
\begin{equation*}
\sup_{i\in I}\rho _{i}(X)=\min \left\{ \gamma \left( X\right) \mid \forall
i\in I,\ \gamma \in \tilde{\Gamma}_{i}\right\} =\min \left\{ \gamma \left(
X\right) \ \left\vert \ \gamma \in \bigcap_{i\in I}\tilde{\Gamma}_{i}\right.
\right\}
\end{equation*}%
also note that ---since the set over which minimization was performed never changed---
the set
\begin{equation*}
\bigcap_{i\in I}\tilde{\Gamma}_{i}\end{equation*}%
consists of all convex risk measures that dominate $\sup_{i\in I}\rho _{i}$.\medskip

\noindent \textbf{Infima.} As well known,%
\begin{equation*}
\inf_{i\in I}\min_{\gamma _{i}\in \Gamma _{i}}\gamma _{i}(X)=\inf_{\gamma
\in \cup _{i\in I}\Gamma _{i}}\gamma (X).
\end{equation*}%
Moreover, if the infimum on the left-hand side is a minimum, then the
infimum on the right-hand side is also a minimum.\medskip

\noindent \textbf{Inf-convolutions.} First observe that for each $%
\boldsymbol{\gamma }=\left( \gamma _{i}\right) _{i\in I}\in
\prod\nolimits_{i\in I}\Gamma _{i}=\boldsymbol{\Gamma }$, and all $%
Y_{1},Y_{2},...,Y_{n}\in \mathcal{X}$ such that $\sum_{i\in I}Y_{i}=0$, we
have%
\begin{equation*}
\sum_{i\in I}\gamma _{i}\left( Y_{i}\right) \geq \sum_{i\in I}\rho
_{i}\left( Y_{i}\right) \geq 0
\end{equation*}%
because $\gamma _{i}\geqq \rho _{i}$, thus each $\square _{i\in I}\gamma
_{i}\in \square _{i\in I}\Gamma _{i}$ is a (well defined) convex risk
measure.

We want to show that:

\begin{itemize}
\item each $\square _{i\in I}\gamma _{i}\in \square _{i\in I}\Gamma _{i}$
dominates $\rho _{\diamond }=\square _{i\in I}\rho _{i}$;

\item $\rho _{\diamond }\left( X\right) =\inf \left\{ \square _{i\in
I}\gamma _{i}\left( X\right) \mid \boldsymbol{\gamma }\in \boldsymbol{\Gamma 
}\right\} =\inf \left\{ \delta \left( X\right) \mid \delta \in \square
_{i\in I}\Gamma _{i}\right\} $.
\end{itemize}

As to the first point, note that, for all $\boldsymbol{\gamma }\in 
\boldsymbol{\Gamma }$ and all $Y_{1},...,Y_{n}\in \mathcal{X}$ such that $%
\sum_{i\in I}Y_{i}=X$, we have%
\begin{equation*}
\sum_{i\in I}\gamma _{i}\left( Y_{i}\right) \geq \sum_{i\in I}\rho
_{i}\left( Y_{i}\right)
\end{equation*}%
but then%
\begin{eqnarray*}
\square _{i\in I}\gamma _{i}\left( X\right) &=&\inf \left\{ \sum_{i\in
I}\gamma _{i}\left( Y_{i}\right) \ \left\vert \ Y_{i}\in \mathcal{X}\text{
for all }i\in I\text{ and\ }\sum_{i\in I}Y_{i}=X\right. \right\} \\
&\geq &\inf \left\{ \sum_{i\in I}\rho _{i}\left( Y_{i}\right) \ \left\vert \
Y_{i}\in \mathcal{X}\text{ for all }i\in I\text{ and\ }\sum_{i\in
I}Y_{i}=X\right. \right\} \\
&=&\rho _{\diamond }\left( X\right)
\end{eqnarray*}%
as wanted.

As to the second, for each $X\in \mathcal{X}$ and all $\varepsilon >0$,
there exist $Y_{1},...,Y_{n}\in \mathcal{X}$ such that $\sum_{i\in I}Y_{i}=X$
and 
\begin{equation*}
\rho _{\diamond }\left( X\right) +\varepsilon \geq \sum_{i\in I}\rho
_{i}\left( Y_{i}\right) =\sum_{i\in I}\bar{\gamma}_{i}\left( Y_{i}\right)
\end{equation*}%
for some $\left( \bar{\gamma}_{1},...,\bar{\gamma}_{n}\right) \in
\prod\nolimits_{i\in I}\Gamma _{i}$ 
because minima are attained in (\ref{eq:op0}); otherwise one could use the
fact that $\sum_{i\in I}\rho _{i}\left( Y_{i}\right) \geq \sum_{i\in
I}\left( \bar{\gamma}_{i}\left( Y_{i}\right) -\varepsilon /2n\right) $. It
follows that 
\begin{equation*}
\rho _{\diamond }\left( X\right) +\varepsilon \geq \square _{i\in I}\bar{
\gamma}_{i}\left( X\right) \geq \inf \left\{ \square _{i\in I}\gamma
_{i}\left( X\right) \mid \left( \gamma _{i}\right) _{i\in I}\in
\prod\nolimits_{i\in I}\Gamma _{i}\right\} =\inf \left\{ \delta \left(
X\right) \mid \delta \in \square _{i\in I}\Gamma _{i}\right\}.
\end{equation*}%
Letting $\varepsilon \rightarrow 0$, we conclude that%
\begin{equation*}
\rho _{\diamond }\left( X\right) \geq \inf \left\{ \delta \left( X\right)
\mid \delta \in \square _{i\in I}\Gamma _{i}\right\}
\end{equation*}%
and the converse inequality descends from the previous point.

Finally, if 
\begin{equation*}
\rho _{\diamond }(X)=\min \left\{ \sum_{i\in I}\rho _{i}\left( Y_{i}\right)
\ \left\vert \ Y_{i}\in \mathcal{X}\text{ for all }i\in I\text{ and\ }
\sum_{i\in I}Y_{i}=X\right. \right\}.
\end{equation*}%
Then, for each $X\in \mathcal{X}$, take $Y_{i}\in \mathcal{X}$ ($i\in I$)\
such that\ $\sum_{i\in I}Y_{i}=X$ and 
\begin{equation*}
\rho _{\diamond }(X)=\sum_{i\in I}\rho _{i}\left( Y_{i}\right).
\end{equation*}%
Since minima are attained in (\ref{eq:op0}), then there exists some $\left( 
\bar{\gamma}_{1},...,\bar{\gamma}_{n}\right) \in \prod\nolimits_{i\in
I}\Gamma _{i}$ such that 
\begin{equation*}
\rho _{\diamond }\left( X\right) =\sum_{i\in I}\rho _{i}\left( Y_{i}\right)
=\sum_{i\in I}\bar{\gamma}_{i}\left( Y_{i}\right)
\end{equation*}%
but then%
\begin{equation*}
\inf \left\{ \square _{i\in I}\gamma _{i}\left( X\right) \mid \left( \gamma
_{i}\right) _{i\in I}\in \prod\nolimits_{i\in I}\Gamma _{i}\right\} =\rho
_{\diamond }\left( X\right) =\sum_{i\in I}\bar{\gamma}_{i}\left(
Y_{i}\right) \geq \square _{i\in I}\bar{\gamma}_{i}\left( X\right)
\end{equation*}%
thus the infimum is attained, which concludes the proof.\hfill $\blacksquare
\bigskip$

\noindent \textbf{Proof of Proposition \ref{coro:robust}.}
We have shown in the proof of Theorem \ref{thm:conv} that  the star-shaped risk measure $\rho$ has representation
 $$
\rho (X)= \min_{Y\in \X} \rho_{\mathcal A_Y} (X) \qquad X\in \mathcal{X},
 $$
where $\rho_{\mathcal A_Y}$ has acceptance set 
$$
\mathcal A_Y  = \mathrm{co}\{Y-\rho (Y) ,0\} -\mathcal{X}^{+} =\{Z\in \X: Z\leqq \alpha(Y-\rho(Y)) \mbox{~for some $\alpha \in[0,1]$}\}.
$$
To get  the representation \eqref{eq:rep-FS}, it suffices to show that each $\rho_{\mathcal A_Y}$ has a representation $\rho_{\mathcal A_Y}=\sup_{Q\in \mathcal{P}}\left\{ \mathbb{E}_{Q}-\alpha _{\gamma }\left( Q\right) \right\}$. This is equivalent to the Fatou property of $\rho_{\mathcal A_Y}$ as in Theorem 4.33 of \cite{FS16}.
The Fatou property of $\rho_{\mathcal A_Y}$ can be stated via the Fatou closure of $\mathcal A_Y$, that is,
$$
X_n \buildrel \mathrm{a.s.} \over \longrightarrow  X,~(X_n)_{n\in \N} \mbox{~is uniformly bounded, and } (X_n)_{n\in \N}  \subseteq \mathcal A_Y ~\Longrightarrow ~X\in \mathcal A_Y,
$$
where $\buildrel \mathrm{a.s.} \over \longrightarrow$ represents almost sure convergence. We next show this Fatou closure. 
For each $n\in \N$, since $X_n\in \mathcal A_Y$, we have $X_n \leqq \alpha_n (Y-\rho(Y))$ for some $\alpha_n\in [0,1]$.
Noting that $[0,1]$ is compact, take a subsequence $(\alpha_{n_k})_{k\in \N}$ of $(\alpha_n)_{n\in \N}$ which converges to a limit, which is denoted by $\alpha_0\in [0,1]$.
We have 
$
X_{n_k} \buildrel \mathrm{a.s.} \over \longrightarrow  X,
$
and therefore
$$
\alpha_0(Y-\rho(Y))= \lim_{k\to \infty}\alpha_{n_k}  (Y-\rho(Y)) \geqq \lim_{k\to \infty}X_{n_k} = X.$$ 
Thus, $X\in \mathcal A_Y$, showing that $\mathcal A_Y$ is Fatou closed. This leads to \eqref{eq:rep-FS}. \hfill $\blacksquare \bigskip $

\noindent \textbf{Proof of Proposition \ref{prop:opt}.} The proof is a
routine verification once the reader recalls that $\rho _{i}=\min_{\gamma
\in \Gamma _{i}}\gamma $ for each $i\in I$ (Theorem \ref{thm:conv}) and $%
\sup_{i\in I}\rho _{i}=\rho _{\vee }=\min_{\gamma \in \tilde{\Gamma}
_{I}}\gamma $ (Theorem \ref{th:rep-opre}).\hfill $\blacksquare \bigskip $

\noindent \textbf{Proof of Theorem \ref{thm:MaoWang}.}  
By Theorem 3.1 of \cite{MW20}, for an $\mathcal{SSD}$-consistent risk measure $\rho$,  the representation \eqref{eq:es1} holds with  
$$
\mathcal G =\{g_Y:(0,1)\to \mathbb{R},~\alpha \mapsto \mathrm{ES}_\alpha(Y)\mid Y\in \mathcal A_\rho \}. 
$$
 The closed interval $[0,1]$ is used in \cite{MW20}, but it is easy to check that  one can also use the open interval  $(0,1)$, since $\alpha\mapsto \mathrm{ES}_\alpha(Y)$ is continuous. 
 
\underline{(i) implies (ii)}. It suffices to check that  the set $\mathcal G$ is   star-shaped when $\rho$ is star-shaped.
This follows immediately from the fact that  $Y\in \mathcal A_\rho$ implies $\lambda Y\in \mathcal A_\rho$ for $\lambda \in [0,1]$, and positive homogeneity of ES gives $\lambda g_Y\in \mathcal G$.

\underline{(ii) implies (i)}. It suffices to check that $\rho$ is star-shaped when $\mathcal G$ is star-shaped. For $\lambda \in [0,1]$,  
noting that $\lambda \mathcal G \subseteq \mathcal G$, 
we have, for $X\in \mathcal X$,
\begin{align*}
\rho (\lambda X)&=\inf_{g\in \mathcal{G}}\sup_{\alpha \in (0,1)}\{\mathrm{ES}_{\alpha
}(\lambda X)-g(\alpha )\}  \\
&\le \inf_{g\in \lambda \mathcal{G}}\sup_{\alpha \in (0,1)}\{\lambda \mathrm{ES}_{\alpha
}( X)-g(\alpha )\} 
 \\
&= \lambda \inf_{g\in  \mathcal{G}}\sup_{\alpha \in (0,1)}\{\mathrm{ES}_{\alpha
}(  X)-g(\alpha )\}   =\lambda \rho(X).
\end{align*}
Hence, $\rho$ is star-shaped. 

The final part of the statement regarding inf-convolutions follows from
Theorem 4.1 of \cite{MW20}. The rest is straightforward.
\hfill $\blacksquare \bigskip $

\noindent \textbf{Proof of Theorem \ref{prop:var}.} \underline{(i) implies (ii)}. We
denote by $X\succcurlyeq _{\mathcal{FSD}}Y$ the fact that $X\in \mathcal{X}$
first-order stochastically dominates $Y\in \mathcal{X}$, which means%
\begin{equation}\label{eq:r1-1}
F_{X}\geq F_{Y},\mbox{~or equivalently,~} \mathrm{VaR}_\alpha(X) \le \mathrm{VaR}_\alpha(Y) \mbox{~for all $\alpha \in (0,1)$},
\end{equation}%
where $F_{X}$ and $F_{Y}$ are the distribution functions of $X$ and $Y$
under $P$. Since $\rho $ is law-invariant and monotonic, 
\begin{equation*}
X\succcurlyeq _{\mathcal{FSD}}Y\implies \rho \left( X\right) \leq \rho
\left( Y\right) .
\end{equation*}

Let $\mathcal{A}_{\rho }=\{Y\in \mathcal{X}:\rho (Y)\leq 0\}$ be the
acceptance set of $\rho $. For each $Y\in \mathcal{A}_{\rho }$, let $%
g_{Y}:\left( 0,1\right) \rightarrow \mathbb{R}$ be given by $g_{Y}(\alpha )=%
\mathrm{VaR}_{\alpha }(Y)$. Note that if $Y\in \mathcal{A}_{\rho }$, then:

\begin{itemize}
\item The same is true for $\beta Y$ for any $\beta \in \lbrack 0,1]$,
hence, by positive homogeneity of \textrm{VaR}, 
\begin{equation*}
\mathcal{G}=\{g_{Z}:Z\in \mathcal{A}_{\rho }\}
\end{equation*}
is a star-shaped set.

\item We have $g_{Y}(0+)\leq 0$. To see this, as well known (see, e.g., \cite[Remark 4.50]{FS16}), $g_{Y}(0+)$ is the essential infimum of $Y$, if it were strictly positive, it would follow that 
$Y\geqq \varepsilon $ for some $\varepsilon >0$, and so $\varepsilon =\rho
\left( \varepsilon \right) \leq \rho \left( Y\right) \leq 0$ because $Y\in 
\mathcal{A}_{\rho }$, a contradiction. 

\item The same is true for any $Z\in \mathcal{X}$ such that $Z\succcurlyeq
_{ \mathcal{FSD}}Y$,  and hence 
\begin{equation*}
\mathcal{A}_{\rho }=\bigcup_{Y\in \mathcal{A}_{\rho }}\{Z\in \mathcal{X}\mid
Z\succcurlyeq _{\mathcal{FSD}}Y\}.
\end{equation*}
\end{itemize}

It follows that 
\begin{align*}
\rho (X)& =\inf \{m\in \mathbb{R}\mid X-m\in \mathcal{A}_{\rho }\} \\
& =\inf \left\{ m\in \mathbb{R}\mid X-m\in \bigcup_{Y\in \mathcal{A}_{\rho
}}\{Z\in \mathcal{X}\mid  Z\succcurlyeq _{\mathcal{FSD}}Y\}\right\} \\
& =\inf \bigcup_{Y\in \mathcal{A}_{\rho }}\left\{ m\in \mathbb{R}\mid X-m\in
\{Z\in \mathcal{X}\mid Z\succcurlyeq _{\mathcal{FSD}}Y\}\right\} \\
& =\inf \bigcup_{Y\in \mathcal{A}_{\rho }}\{m\in \mathbb{R}\mid  X-m\succcurlyeq
_{\mathcal{FSD}}Y\} \\
& =\inf_{Y\in \mathcal{A}_{\rho }}\inf \{m\in \mathbb{R}\mid  X-m\succcurlyeq _{ 
\mathcal{FSD}}Y\}.
\end{align*}%
By \eqref{eq:r1-1}, we have
\begin{align*}
 \inf \{m\in \mathbb{R}\mid  X-m\succcurlyeq _{ 
\mathcal{FSD}}Y\}  
&=  \inf \{m\in \mathbb{R}\mid  \mathrm{VaR}
_{\alpha }(X  -m) \le \mathrm{VaR}
_{\alpha }(Y) \mbox{~for all $\alpha \in (0,1)$}\}  
\\&=  \inf \{m\in \mathbb{R}\mid m \ge    \mathrm{VaR}
_{\alpha }(X) -  \mathrm{VaR}
_{\alpha }(Y) \mbox{~for all $\alpha \in (0,1)$}\}  
\\ &= \sup_{\alpha \in (0,1)}\{\mathrm{VaR}
_{\alpha }(X)-\mathrm{VaR}_{\alpha }(Y)\}.
\end{align*}%
Therefore, 
\begin{align*} 
\rho (X)&=\inf_{Y\in \mathcal{A}_{\rho }}\inf \{m\in \mathbb{R}\mid  X-m\succcurlyeq _{ 
\mathcal{FSD}}Y\}\\& =\inf_{Y\in \mathcal{A}_{\rho }}\sup_{\alpha \in (0,1)}\{\mathrm{VaR}
_{\alpha }(X)-\mathrm{VaR}_{\alpha }(Y)\} \\
& =\inf_{Y\in \mathcal{A}_{\rho }}\sup_{\alpha \in (0,1)}\{\mathrm{VaR}
_{\alpha }(X)-g_{Y}(\alpha )\}   =\inf_{g\in \mathcal{G}}\sup_{\alpha \in (0,1)}\{\mathrm{VaR}_{\alpha
}(X)-g(\alpha )\},
\end{align*}
thus showing (\ref{eq:var1}).

\underline{(ii) implies (i)}. Monotonicity, translation invariance, and law-invariance
of $\rho $ are obvious. Moreover, 
\begin{equation*}
\rho (0)=\inf_{g\in \mathcal{G}}\sup_{\alpha \in (0,1)}\{-g(\alpha
)\}=\inf_{g\in \mathcal{G}}-g(0+)=-\sup_{g\in \mathcal{G}}g(0+)=0
\end{equation*}%
because star shapedness of $\mathcal{G}$ guarantees $0\in \mathcal{G}$ and $%
g(0+)\leq 0$ for all $g\in \mathcal{G}$, and hence $\rho $ is normalized.
This together with monotonicity and translation invariance also shows that $%
\rho $ takes real values. Hence $\rho $ is a law-invariant risk measure.
Finally, if $\lambda >1$ and $X\in \mathcal{X}$, then%
\begin{align*}
\rho (\lambda X)& =\inf_{g\in \mathcal{G}}\sup_{\alpha \in (0,1)}\{\mathrm{\
VaR}_{\alpha }(\lambda X)-g(\alpha )\}=\inf_{g\in \mathcal{G}}\sup_{\alpha
\in (0,1)}\{\lambda \mathrm{VaR}_{\alpha }(X)-g(\alpha )\} \\
& =\lambda \inf_{g\in \mathcal{G}}\sup_{\alpha \in (0,1)}\left\{ \mathrm{VaR}
_{\alpha }(X)-\frac{g(\alpha )}{\lambda }\right\} =\lambda \inf_{h\in
\lambda ^{-1}\mathcal{G}}\sup_{\alpha \in (0,1)}\left\{ \mathrm{VaR}_{\alpha
}(X)-h(\alpha )\right\} \\
& \geq \lambda \inf_{g\in \mathcal{G}}\sup_{\alpha \in (0,1)}\{\mathrm{VaR}
_{\alpha }(X)-g(\alpha )\}=\lambda \rho (X)
\end{align*}%
because $\lambda ^{-1}\mathcal{G}\subseteq \mathcal{G}$. Therefore, $\rho $
is star-shaped.

The final part of the statement regarding inf-convolutions follows from
Theorem 2 of Liu, Wang, and Wei \cite{LWW20}. The rest is straightforward.\hfill $%
\blacksquare \bigskip $

\noindent \textbf{Proof of Proposition \ref{prop:r1-1}.} 
 Suppose that $\rho$ is an $\mathcal{SSD}$-consistent risk measure dominating $\mathrm{VaR}_\alpha$. We will show   $\rho \geqq \mathrm{ES}_\alpha$. Suppose otherwise. By translation invariance, there exists $X\in \mathcal X$ such that $\rho(X)\le 0$ and $\mathrm{ES}_\alpha(X)>0$. 
Since $\alpha\mapsto \mathrm{ES}_\alpha(X)$ is continuous, there exists $ \alpha'\in (0,\alpha)$ such that $\mathrm{ES}_{\alpha'}(X)>0$, and denote by $x= \mathrm{ES}_{\alpha'}(X)$. 
Construct a random variable $Y$ by 
$$
Y= X\id_{A^c} + x \id_{A},
$$
where $A$ is an $\alpha'$-tail event of $X$ as defined by \cite{WZ21}, meaning that $P(A)=1-\alpha'$
and $X(\omega) \ge X(\omega')$ for almost every $\omega\in A$ and $\omega'\in A^c$.
Note that $X\le 0$ almost surely on $A^c$ because $\mathrm{VaR}_{\alpha'}(X) \le \mathrm{VaR}_\alpha(X)\le \rho(X) \le0$.
Moreover, since $x=\mathrm{ES}_{\alpha'}(X) = \E_P[X|A]$,
we have $Y=\E_P[X|Y]$. By Jensen's inequality, we have $Y\succcurlyeq_{\mathcal{SSD}} X$,
and $\mathcal{SSD}$-consistency of $\rho$ gives $\rho(Y)\le \rho(X) $.
However, since $P(Y>0)=P(A) = 1-\alpha' > 1-\alpha$, we have $\mathrm{VaR}_\alpha(Y)>0$,
which leads to $\rho(Y) <\mathrm{VaR}_\alpha(Y)$, a contradiction.  
Therefore, we obtain  $\rho \geqq  \mathrm{ES}_\alpha$.  \hfill $%
\blacksquare \bigskip $

\end{APPENDIX}
\theendnotes

\bigskip


\textit{In memoriam.} I met Erio Castagnoli (Mantova, July 2nd, 1943 --- January 9th, 2019), in September, a quarter century ago, at the conference in Urbino where I presented my first paper. Erio gave a lecture on the importance of the Bipolar Theorem of Functional Analysis in Decision Theory: I was hypnotized by the beauty. I knew a bit of the first topic, nothing of the second. We started to work on it that very afternoon.

Erio changed my life, and I will owe him forever.\bigskip

\hfill Fabio Maccheroni  

\end{document}